\pdfoutput=1
\documentclass[a4paper,20pt]{article}
\usepackage{multirow}
\usepackage{amsmath}
\usepackage{mathptmx}
\usepackage{xcolor}
\usepackage{amsfonts}
\usepackage{amssymb}
\usepackage{amstext}
\usepackage{mathtools}
\usepackage{import,breqn}
\usepackage{graphicx}
\usepackage{caption,subcaption}
\usepackage{float}
\usepackage{bm}
\usepackage{makecell}
\usepackage{authblk}
\usepackage{hyperref}
\usepackage{orcidlink}
\usepackage{cite}
\usepackage{epsfig}%
\usepackage{pdfrender}
\usepackage{cancel}
\usepackage{tabularx}
\usepackage{adjustbox}

\hypersetup{
	colorlinks=true,
	linkcolor=blue,
	urlcolor=blue,
	citecolor=red
}
\title{Illuminating the homogeneous and isotropic cosmological solution space of $f(R)$ and $f(Q)$ gravity with a unified dynamical systems approach}

\author{Saikat Chakraborty\thanks{Electronic address: \texttt{saikatchakraborty@duytan.edu.vn}}}
\affil{Institute of Research and Development, Duy Tan University, Da Nang 550000, Vietnam}
\affil{Faculty of Natural Sciences, Duy Tan University, Da Nang 550000, Vietnam}
\date{}

\begin{document}

\maketitle

\begin{abstract}
Going beyond the traditional dynamical system framework constructed in terms of the Hubble-normalized dimensionless dynamical variables, we formalize a unified dynamical system framework for $f(R)$ gravity and the coincident gauge $f(Q)$ gravity in terms of three distinct set of dimensionless dynamical variables: (1) the \emph{kinematical} cosmographic parameters satisfying a recursion relation, (2) the set consisting of the dark energy equation of state parameter $w$ and its derivatives with respect to $N=\ln a$, and the standard matter density abundance parameter $\Omega_m$, and (3) a set of \emph{theory space} parameters $\{m_i\}$ satisfying a recursion relation. Our formulation is free of any auxiliary variable, and for any given problem, one needs only a finite number of variables from two of the three sets above to obtain a closed system. We show that the resulting dynamical system can be closed in three equivalent ways: (1) by specifying a given theory, (2) by requiring that the universe follows a certain kind of evolution, or (3) by requiring that the dark energy equation of state follows a certain kind of evolution. The choice of these variables helps one to distinguish the kinematics and dynamics of the universe, as well as the dark energy equation of state evolution.

We illustrate theory closure with the completely generic Hu-Sawicki $f(R)$ family without any assumption on its parameters, and the $f(Q)=-2\Lambda+Q+\beta\sqrt{-Q}$ model. For the generic Hu-Sawicki family, we identify a constraint surface on the theory space, on which any cosmological solution of the theory must reside. In the $f(Q)$ example, a broad set of initial conditions gives a heteroclinic connection between a GR-matter-dominated epoch and a late-time de Sitter epoch. The cosmographic closure approach and the equation-of-state closure approach are applied for a comparative study of the $f(R)$ and $f(Q)$ gravity that are \emph{kinematically} equivalent to $\Lambda$CDM (reproducing $j(z)=1$; $j$ being the jerk parameter) and \emph{dynamically} equivalent to $\Lambda$CDM (reproducing $w(z)=-1$), respectively. In both studies, it is found that GR behaves like a cosmological past attractor for $f(Q)$ gravity, but not for $f(R)$ gravity.
\end{abstract}

\clearpage 

\tableofcontents

\clearpage

\section{Introduction}

Cosmological model building for the observed late-time accelerated expansion of the universe remains one of the central problems of current theoretical cosmology. While the General Relativistic (GR) $\Lambda$CDM model accommodates this acceleration through a cosmological constant $\Lambda$, its well-documented theoretical difficulties, together with recent observational hints from the DESI collaboration's Baryon Acoustic Oscillation measurements about the possible deviation of the dark energy equation of state from $w=-1$ and its time evolution \cite{DESI:2025zgx}, continue to motivate both alternative explanations of cosmic acceleration through modifications of GR and more flexible, model-independent parametrizations of the dark energy sector \cite{papantonopoulos2015modifications,saridakis2021modified,Copeland:2006wr}. Since we are exclusively interested in this late-time regime throughout the paper, we ignore the fractional contribution from radiation and global spatial curvature in the total energy budget of the universe for this particular work ($|\Omega_r|,|\Omega_K|\ll1$).

Among the many modified gravity theories proposed in this context, $f(R)$ gravity \cite{Sotiriou:2008rp,DeFelice:2010aj} and $f(Q)$ gravity \cite{BeltranJimenez:2019tme} occupy a particularly prominent place, owing respectively to their conceptual proximity to General Relativity and to the comparative algebraic simplicity of their field equations. Both have been extensively applied to late-time cosmology, in particular to explore whether the extra geometric degree of freedom they introduce -- which is curvature-based in the former case, and non-metricity-based in the latter -- can drive  the observed cosmic acceleration. Here, at the very onset, we note that there are actually three inequivalent symmetric teleparallel connection branches compatible with global spatial flatness, spatial homogeneity and isotropy \cite{Hohmann:2021ast}. In this paper we work exclusively with the first of these branches, for which the coincident gauge coincides with the comoving coordinates in which the FLRW metric is conventionally written \cite{Hohmann:2021ast}. This branch is commonly referred to as the coincident gauge of $f(Q)$ cosmology, and whenever we refer to $f(Q)$ gravity in what follows, this is the branch we have in mind.

The dynamical system approach has, for decades, been one of the most powerful and popular tools for investigating the late-time cosmological viability of such theories, since it characterizes the qualitative behaviour of a cosmological model -- its fixed points, their stability, and the possible heteroclinic connections between them -- without requiring an explicit solution of the underlying field equations \cite{wainwright2009dynamical,coley2003dynamical,Bahamonde:2017ize}. The traditional dynamical system formulations of cosmology in $f(R)$ gravity \cite{Amendola:2006we,Carloni:2007br,Chakraborty:2021mcf} and $f(Q)$ gravity \cite{Dutta:2025fqw,Khyllep:2026pku}, which are built on Hubble-normalized dimensionless dynamical variables, suffer from an important limitation. These formulations commonly eliminate the curvature or nonmetricity scalar between $r=Xf_X/f$ and $m=Xf_{XX}/f_X$ ($X=R,Q$). A single chart of this form requires a single-valued relation $m=m(r)$. Where $r(X)$ turns, distinct branches can give different values of $m$ at the same $r$, as we discuss in Secs.~\ref{sec:HS} and \ref{sec:sqrt_f(Q)}.

A second, more conceptual limitation of the traditional Hubble-normalized formulation is that it mixes up the \emph{kinematics} and \emph{dynamics} of the universe. For example, in the traditional dynamical system formulation of $f(R)$ gravity \cite{Amendola:2006we,Carloni:2007br} as well as $f(Q)$ gravity \cite{Dutta:2025fqw}, an important dimensionless dynamical variable is
\begin{equation}
    \frac{f_{XX}\dot{X}}{Hf_X}=\frac{Xf_{XX}}{f_X}\cdot\frac{\dot{X}}{HX}=m\cdot\frac{\dot{X}}{HX}\,.  \nonumber
\end{equation}
For $X=R=6H^2(1-q)$ or $X=Q=-6H^2$\,\footnote{We are following the convention of \cite{Dutta:2025fqw}. It is to be noted that in some different convention, $Q=6H^2$ \cite{BeltranJimenez:2019tme}.}, the quantity $\dot{X}/(HX)$ is purely \emph{kinematic} and expressible entirely in terms of the cosmographic quantities, whereas the quantity $m$ is clearly \emph{dynamic} and characterizes the deviation from GR ($f(X)=-2\Lambda+X$). The two are nevertheless entangled together within a single dynamical variable, obscuring which part of the resulting phase space behaviour should be attributed to the kinematics of the expansion and which to the underlying gravitational dynamics.

Moreover, in the traditional formulation the dynamical variables are, by and large, not the dimensionless quantities directly constrained by observational data. For example, an important dimensionless dynamical variable in the traditional formulation is $\Omega=\frac{8\pi G}{f_X}=\frac{\Omega_m}{f_X}$, whereas the quantity actually constrained by data is $\Omega_m$ itself. The entanglement of cosmographic parameters within the definition of $\frac{f_{XX}\dot{X}}{Hf_X}$, noted above, is another instance of the same issue. Together, these limitations make it considerably harder to connect the qualitative phase space dynamics of a given model with actual cosmological phenomenology.

These limitations are precisely what motivate us to reframe the entire dynamical system formulation for $f(R)$ and $f(Q)$ gravity in the present work. In carrying this out, we in effect formalize the dynamical system framework for these two very important families of gravity theories, which can be utilized to make a comparative analysis of them from the cosmological perspective. As a key element of this process, in parallel to the standard cosmographic hierarchy, we introduce a \emph{theory-space} hierarchy, which we term the $\{m_i\}$-hierarchy, characterizing the ``shape'' of the theory function $f$ order by order. Our dynamical system formulation, built around these two hierarchies, offers three distinct advantages over the traditional Hubble-normalized formulation: (1) there is no issue of auxiliary variables that need to be expressed in terms of other dynamical variables -- every variable appearing in our formulation is genuinely dynamical in its own right; (2) the kinematics and the dynamics of the universe are cleanly separated from the very outset; and (3) quantities that are directly constrained by observational data -- $\Omega_m$, the cosmographic parameters, and the dark energy equation of state parameter together with its derivatives -- feature as phase space variables in their own right, making it considerably easier to connect the resulting phase space with observational phenomenology.

A central consequence of this reformulation is that a dynamical system can be closed, and thereby rendered finite-dimensional, in three logically distinct ways: (1) by specifying a given theory function $f(R)$ or $f(Q)$; (2) by requiring that the universe evolve in a specific manner, encoded in an algebraic cosmographic relation among the cosmographic parameters; or (3) by requiring that the dark energy equation of state parameter evolve in a specific manner, encoded in an algebraic relation involving $w$ and its derivatives. Correspondingly, there are three distinct sets of dimensionless dynamical variables at our disposal: (1) the theory-space parameters $\{m_i\}$; (2) the cosmographic parameters; and (3) the matter density parameter $\Omega_m$ plus the dark energy equation of state parameter and its derivatives. For any given problem, a closed dynamical system can be constructed using a finite number of parameters drawn from either set (1) \& set (2), or set (1) \& set (3) -- according to which of the three closure strategies above is adopted. In each case, the Friedmann constraint serves to determine the quantity $\Omega_m/f_X$ ($X=R,Q$).

The remainder of the paper is organized as follows. In Sec.~\ref{sec:hierarchy} we introduce the cosmographic and theory-space hierarchies together with their recursion relations, which form the mathematical backbone of the entire formalism. Sec.~\ref{sec:closure_strategy} lays out the general closure strategy sketched above in more detail. Secs.~\ref{sec:f(R)} and \ref{sec:f(Q)} then implement this strategy explicitly for $f(R)$ and $f(Q)$ gravity respectively, presenting the dynamical system corresponding to each of the three closure choices. Secs.~\ref{sec:HS} and \ref{sec:sqrt_f(Q)} illustrate theory closure for the completely generic Hu-Sawicki $f(R)$ family and the $f(Q)=-2\Lambda+Q+\beta\sqrt{-Q}$ model, respectively. Secs.~\ref{sec:LCDM_mimicking} and \ref{sec:cosm_const_mimicking} illustrate, respectively, the cosmographic-closure and equation-of-state-closure branches with the simplest possible worked example in each case: $f(R)$ and $f(Q)$ theories that are kinematically ($j=1$) and dynamically ($w=-1$) equivalent to $\Lambda$CDM. In Sec.~\ref{sec:kin_vs_dyn} we show how the phase spaces of the theories kinematically and dynamically equivalent to $\Lambda$CDM are related, which brings out a structural feature of the formalism, and provides a geometric insight into the dark degeneracy. Finally, we summarize our results and discuss avenues for future work in Sec.~\ref{sec:summary}. Appendices \ref{app:HS_constraint_surface} and \ref{app:HS_parameter_determination} lists some long expressions relevant to the generic Hu-Sawicki case.

\section{The two hierarchies and recursion relations}\label{sec:hierarchy}

We start with the well-known cosmographic series expansion characterizing the evolution of the universe \cite{Dunsby:2015ers,Bolotin:2018xtq}
\begin{equation}
    a(t) = a_0 \left[1 + H_0(t-t_0) - \frac{1}{2!}q_0 H_0^2(t-t_0)^2 + \frac{1}{3!}j_0 H_0^3(t-t_0)^3 + ......\right]\,.
\end{equation}
From the series \eqref{cosm_series} one can also calculate
\begin{align}\label{cosm_series}
    H(t) &= H_0\left[1 - (1+q_0) H_0(t-t_0) + \frac{1}{2!}(2+j_0+3q_0) H_0^3(t-t_0)^2 + ......\right]\,.
\end{align}
The above expansion gives rise to a hierarchy of dimensionless kinematic parameters 
\begin{align}\label{cosm_hierarchy}
    c_0 = \frac{\ddot{a}}{aH^2} =: -q\,,
    \qquad
    c_1 = \frac{\dddot{a}}{aH^3} =: j\,,
    \qquad
    c_2 = \frac{a^{(4)}}{aH^4} \equiv s\,,\,....
\end{align}
which one can generalise to arbitrary order by setting $c_n = a^{(n+2)}/(aH^{n+2})$. Here, the index in round brackets stands for the order of differentiation $a^{(n)} = d^na/dt^n$.

The kinematic parameters satisfy the differential recursion relation
\begin{equation}
    \label{cosm_recursion}
    c_i' = c_{i+1} - \left[1-(i+2)(1-c_0)\right]c_i\,,
\end{equation}
and the prime stands here for differentiation with respect to $N=\log a$, i.e., $c_i' = dc_i/(Hdt)$.

The first terms in the recursion relation~\eqref{cosm_recursion} give
\begin{subequations}
    \label{CP_rel}
    \begin{eqnarray}
        j &=& 2q^{2} + q - \frac{1}{H}\frac{dq}{dt}\,,
        \\
        s &=& \frac{1}{H}\frac{dj}{dt} - j(2 + 3q)\,,\label{s}
        \\
        l &=& \frac{1}{H}\frac{ds}{dt} - s(3+4q)\,.\label{l}
    \end{eqnarray}
\end{subequations}

Many modified theories of gravity contain an unknown function $f(X)$ which is chosen using either a heuristic approach or a mixed argument based on naturalness and simplicity. Here, $X$ stands for the main quantity describing the particular model. In the main body of this paper we will consider two cases: $X=R$ (Ricci curvature scalar) corresponding to $f(R)$ gravity and $X=Q$ (nonmetricity scalar) corresponding to $f(Q)$ gravity. In order to construct a unified formalism, let us begin with a series expansion of $f(X)$ around an arbitrary value $X_0$ where $f$ is assumed to be analytic:
\begin{align}
    f(X) = f(X_0) + f'(X_0)(X-X_0) + \frac{1}{2!}f''(X_0)(X-X_0)^2 +
    \frac{1}{3!}f'''(X_0)(X-X_0)^3 + \ldots \,.
\end{align}
We can rewrite this expansion as
\begin{multline}
    f(X) = f(X_0)\Bigl[ 1+ \frac{f'(X_0) X_0}{f(X_0)}(X/X_0-1) + \frac{1}{2!}\frac{f''(X_0)X_0}{f'(X_0)}\frac{f'(X_0)X_0}{f(X_0)}(X/X_0-1)^2 
    \\+
    \frac{1}{3!}\frac{f'''(X_0) X_0}{f''(X_0)}\frac{f''(X_0) X_0}{f'(X_0)}
    \frac{f'(X_0) X_0}{f(X_0)} + \ldots \Bigr]\,,
    \label{eq:expansion1}
\end{multline}
which motivates the introduction of the dimensionless parameters $m_i$, characterising the ``shape'' of the theory function $f$:
\begin{align}
    \label{m_i_hierarchy}
    m_0 = \frac{Xf_X}{f} = \frac{d\ln f}{d\ln X}\,,\qquad m_1 = \frac{Xf_{XX}}{f_X} = \frac{d\ln f_X}{d\ln X}\,,\qquad m_i = \frac{Xf^{(i+1)}}{f^{(i)}} = \frac{d\ln f^{(i)}}{d\ln X}\,,.....
    \end{align}
This $m_i$ hierarchy characterises the \emph{dynamics} of the model, in the sense that they determine the underlying gravitational theory. Using this notation, we can write the expansion~\eqref{eq:expansion1} as follows
\begin{align}
    f(X) = f(X_0)\Bigl[ 1+ m_0(X/X_0-1) + \frac{1}{2!}m_0m_1(X/X_0-1)^2 +
    \frac{1}{3!}m_0m_1m_2(X/X_0-1)^3 + \ldots \Bigr]\,.
    \label{eq:expansion2}
\end{align}

The parameters $m_i$ follow the recursion relation
\begin{align}
    \label{m_i_recursion}
    \frac{d\ln m_i}{d\ln X} = m_{i+1} - m_i + 1\,.
\end{align}
It can be noted that, at least for a metric theory of gravity, i.e., when the metric is the only independent dynamical quantity in the action, the quantity $\dot{X}/(H X)$ is always expressible in terms of the cosmographic parameters on the FLRW minisuperspace. 

The two sets of recursion relations~\eqref{cosm_recursion} and~\eqref{m_i_recursion} form a closed system of equations. Whereas the \emph{kinematical} cosmographic parameters evolve on their own according to the recursion relation \eqref{cosm_recursion}, the evolution of the \emph{dynamical} theory parameters is coupled to that of the cosmographic parameters, because of the appearance of the $\dot{X}/(HX)$ term in Eq.\eqref{m_i_recursion}. Mathematically speaking, if one is supplied with the explicit global form of the two functions $c_0(t)$ and $m_0(t)$, the coupled system of recursion relations \eqref{cosm_recursion} and \eqref{m_i_recursion} then \emph{uniquely} determines the entire kinematics and dynamics of the cosmological system. Such separation between the \emph{kinematics} and the \emph{dynamics} is at the heart of the dynamical system formulation that we present below.

\section{Formalism and closure strategy}\label{sec:closure_strategy}

The standard statefinder relations are \cite{Sahni:2002fz,Chakraborty:2025rvc,Chakraborty:2026zhr,Worsley:2026ijo}
\begin{subequations}
\begin{eqnarray}
\label{q}
q &=& \frac{1}{2} + \frac{3}{2}w(1-\Omega_m)\,,\\
\label{j}
j &=& 1 + \frac{3}{2}(1-\Omega_m)(3w +3w^2 - w^\prime)\,,\\
\frac{j-1}{3(q-1/2)} &=& 1+w-\frac{w'}{3w}\,,
\label{statefinder_s}
\end{eqnarray}
\end{subequations}
where $\Omega_m=\frac{\kappa\rho}{3h^2},\,w'=\frac{dw}{dN}$, with $w$ is the dark energy equation of state parameter and $N=\ln a$.

In the traditional dynamical system formulation of $f(R)$ gravity \cite{Amendola:2006we,Bahamonde:2017ize}, the quantities $\{r(R),m(R)\}=\left\lbrace\frac{Rf'}{f},\frac{Rf''}{f'}\right\rbrace$ are taken not as dynamical variables but as auxiliary variables. Sometimes, a quantity $\Gamma=\frac{f'}{Rf''}=\frac{1}{m}$ is used instead of $m$ \cite{Carloni:2007br,Chakraborty:2021mcf,Chakraborty:2023clu}, which is just a matter of convention. The quantity $r$ can usually be expressed in terms of the dynamical variables. Provided that, for a given theory $f(R)$, $r=r(R)$ can be inverted to obtain $R=R(r)$, one can express $m=m(r)$. This closes the dynamical system. The traditional dynamical system for $f(Q)$ gravity is formulated the same way \cite{Boehmer:2022wln,Dutta:2025fqw}.

Instead, we take a somewhat more non-traditional approach to construct the dynamical systems here. Note that in the usual traditional dynamical system formulation of $f(R)$ or $f(Q)$ gravity, formulated in terms of the so-called Hubble-normalized dimensionless dynamical variables, the dynamical variables are not directly related to cosmic observables. In other words, they are not the quantities that can be directly constrained from cosmological observations. We, however, utilize, as the dynamical variables, the standard quantities that are directly constrained from data ---- the cosmographic parameters $q,\,j,\,s$, etc., the standard matter density contrast $\Omega_m$, the dark energy equation of state parameter $w$ and its derivatives $w',\,w''$, etc. We also elevate the so-called auxiliary variables $r,\,m$ to the status of dynamical variables of their own right. Moreover, we will also use, whenever required, the higher-order parameters in the $m_i$-hierarchy as additional dynamical variables.

In this way, we can reformulate the dynamical system in a way that clearly distinguishes the two sets of variables spanning its cosmological phase space --- a set of variables that are essentially observable parameters like $q,\,j,\,s,\,w,\,w',\,\Omega_m$, etc. --- and a set of variables that are essentially qualitative parameters describing the ``shape'' of the theory: the $m_i$-parameters. There are multiple motivations behind reformulating the dynamical system this way. By utilizing the standard cosmological parameters as dynamical variables, it becomes much easier to connect the qualitative results from the cosmological phase space with actual cosmological observations. Using the hierarchy of theory space parameters $m_i$ as dynamical variables, as we will see, will extend the applicability of the dynamical system formulation.

A dynamical system can essentially be closed in three different manners: by specifying a given theory function $f(R)$ or $f(Q)$, by specifying a given cosmology, or by specifying a dark energy equation of state parametrization. Therefore, one can either choose to investigate an explicitly specified given theory, or choose to investigate an implicitly specified underlying theory that reproduces a given cosmological evolution or follows a certain dark energy equation of state parametrization. In the case of investigating the cosmological phase space of a given theory, one has the option of choosing the set of dynamical variables that corresponds to the cosmological observable parameters as either $\{q,j\}$ or $\{\Omega_m,w,w'\}$.

The next two sections, Sec.~\ref{sec:f(R)} and Sec.~\ref{sec:f(Q)}, will make our approach clear by implementing them in $f(R)$ and $f(Q)$ gravity respectively.

\section{$f(R)$ gravity}\label{sec:f(R)}

A physically illuminating way to express the cosmological field equations in $f(R)$ gravity, which clearly demarcates the conserved curvature fluid that serves as the potential dark energy, is given by \cite{MacDevette:2024wpg,Chakraborty:2025lkz}
\begin{subequations}\label{fieldeqs_f(R)}
    \begin{align}
        3H^{2} &= \kappa\rho_{\rm tot} = \kappa\rho + \kappa\rho_{\rm DE}\,,
        \\
        -\left(2\dot{H} + 3H^{2}\right) &= \kappa P_{\rm tot} = \kappa P_{\rm DE}\,,
    \end{align}     
\end{subequations}
where
\begin{subequations}\label{curv_fluid}
    \begin{align}
         \kappa\rho_{\rm DE}&=\frac{1}{2}(Rf_R-f)-3H\dot{f_R}+3H^{2}(1-f_R)\,,\label{DE_ed_f(R)}\\
         \kappa P_{\rm DE}&=\ddot{f_R}+2H\dot{f_R}-\frac{1}{2}(Rf_R-f)-(2\dot{H}+3H^2)(1-f_R)\,. \label{DE_p_f(R)}
    \end{align}
\end{subequations}
The dark energy equation of state $w = \frac{P_{\rm DE}}{\rho_{\rm DE}}$ is 
\begin{equation}\label{DE_eos_f(R)}
    w = \frac{\ddot{f_R}+2H\dot{f_R}-\frac{1}{2}(Rf_R-f)-(2\dot{H}+3H^2)(1-f_R)}{\frac{1}{2}(Rf_R-f)-3H\dot{f_R}+3H^{2}(1-f_R)} = \frac{2}{3}\left(\frac{q-1/2}{1-\Omega_m}\right)\,.
\end{equation}
In this form, the non-relativistic matter and the dark energy fluid are separately conserved
\begin{subequations}\label{cons_f(R)}
    \begin{eqnarray}
        && \dot\rho + 3H\rho = 0\,.\\
        && \dot\rho_{\rm DE} + 3H\rho_{\rm DE}(1+w) = 0\,.
    \end{eqnarray}
\end{subequations}

\subsection{Closing by a given theory}\label{subsec:theory_f(R)}

In this subsection, we present the construction of how a dynamical system can be obtained once one is supplied with a given theory $f(R)$. The dynamical system can be written in two equivalent ways. The first way to write a dynamical system is as follows:
\begin{subequations}\label{DS_f(R)_1}
\begin{align}
\frac{\Omega_m}{f_R} &= 1 + (1-q)\left(\frac{1}{r}-1\right) + m\left(\frac{j-q-2}{1-q}\right)\,,\label{constr_f(R)_1}
\\
\frac{dq}{dN} &= 2q^2 + q - j\,,
\\
\frac{dj}{dN} &= -[2(1-q^2)-3q(j-q-2)] + \frac{(1-q)(2-q)}{m} - \frac{3(1-q)^2}{m\,r} - \frac{(j-q-2)^2}{(1-q)}\,m_2
\\
\frac{dr}{dN} &= r(1-r+m)\left(\frac{j-q-2}{1-q}\right)\,,
\\
\frac{dm}{dN} &= m(1-m+m_2)\left(\frac{j-q-2}{1-q}\right)\,,
\\
\frac{dm_2}{dN} &= m_2(1-m_2+m_3)\left(\frac{j-q-2}{1-q}\right)\,,
\\
& .................................... \nonumber
\\
& .................................... \nonumber 
\\
\frac{dm_i}{dN} &= m_i(1-m_i+m_{i+1})\left(\frac{j-q-2}{1-q}\right)\,.
\end{align}
\end{subequations}
Notice that the dynamical system \eqref{DS_f(R)_1} is singular at $q=1$. However, the value $q=1$ is not attained during late-time cosmic evolution, which typically starts at $q=1/2$ during the matter-dominated epoch. 

In late time cosmology, usually we have $R=6H^2(1-q)>0$. Since, by definition $m=\frac{Rf_{RR}}{f_R}$, and the conditions $f_R<0$ and $f_{RR}<0$ is usually associated with ghost instability and Dolgov-Kawasaki instability in late time $f(R)$ cosmology, the physically viable domain of the phase space is given by 
\begin{equation}\label{phys_viab_f(R)_1}
    {\cal D} = \left\lbrace \{q,j,r,m,....,m_i\}\in \mathbb{R}^{i+3}: m\geq0\,\,\wedge\,\,1 + (1-q)\left(\frac{1}{r}-1\right) + m\left(\frac{j-q-2}{1-q}\right)\geq 0 \right\rbrace\,,
\end{equation}
with the last inequality coming from the Friedmann constraint \eqref{constr_f(R)_1}.

It can be noted that the dynamical system \eqref{DS_f(R)_1} is also apparently singular at $m=0$. However, if we confine ourselves strictly within the physically viable region of the phase space where the condition $m>0$ holds (with the singular limit $m\to0$ implying GR), then the dynamical system \eqref{DS_f(R)_1} can be regularized by a time redefinition on the phase space of the form 
\begin{equation}
    dN \to \tilde d\tilde{N} = \frac{dN}{m}\,.
\end{equation}
In terms of the new time variable $\tilde{N}$, the dynamical system \eqref{DS_f(R)_1} can be rewritten as
\begin{subequations}\label{DS_f(R)_2}
\begin{align}
\frac{\Omega_m}{f_R} &= 1 + (1-q)\left(\frac{1}{r}-1\right) + m\left(\frac{j-q-2}{1-q}\right)\,,
\\
\frac{dq}{d\tilde{N}} &= m(2q^2 + q - j)\,,
\\
\frac{dj}{d\tilde{N}} &= -m[2(1-q^2)-3q(j-q-2)] + (1-q)(2-q) - \frac{3(1-q)^2}{r} - \frac{(j-q-2)^2}{(1-q)}\,m\,m_2
\\
\frac{dr}{d\tilde{N}} &= r\,m(1-r+m)\left(\frac{j-q-2}{1-q}\right)\,,
\\
\frac{dm}{d\tilde{N}} &= m^2(1-m+m_2)\left(\frac{j-q-2}{1-q}\right)\,,
\\
\frac{dm_2}{d\tilde{N}} &= m\,m_2(1-m_2+m_3)\left(\frac{j-q-2}{1-q}\right)\,,
\\
& .................................... \nonumber
\\
& .................................... \nonumber 
\\
\frac{dm_i}{d\tilde{N}} &= m\,m_i(1-m_i+m_{i+1})\left(\frac{j-q-2}{1-q}\right)\,.
\end{align}
\end{subequations}
The dynamical system \eqref{DS_f(R)_2} is, by construction, regular at the GR limit $m\to0$.

An alternative way to write a dynamical system is as follows:
\begin{subequations}\label{DS_f(R)_3}
\begin{align}
\frac{\Omega_m}{f_R} &= -\frac{r \left(6 m (w (3 w+2) (\Omega_m-1)+1)+9 w^2 (\Omega_m-1)^2-1\right)-6 m r (\Omega_m-1) w'-(3 w (\Omega_m-1)+1)^2}{2 r (3 w (\Omega_m-1)+1)}\,,\label{constr_f(R)_2}
\\
\frac{d\Omega_m}{dN} &= 3w\Omega_m(1-\Omega_m)\,,
\\
\frac{dw}{dN} &= w'\,,
\\
\frac{dw'}{dN} &= -\frac{1}{2(1-\Omega_m)} \Bigg[ \frac{(w(\Omega_m - 1) + 1)(3w(\Omega_m - 1) + 1)}{m} - \frac{(3w(\Omega_m - 1) + 1)^2}{mr} \nonumber \\
&\qquad - \frac{6 m_2 \left(-(\Omega_m - 1)w' + w(3w + 2)(\Omega_m - 1) + 1\right)^2}{1-3w(1-\Omega_m)} - 5 \Bigg] \nonumber \\
&\qquad - \frac{3}{2}w(3w^2 + 2w - w')(1-\Omega_m) - \frac{1}{2}(18w^3 + 27w^2 - 18ww' - 9w' + w)\,,
\\
\frac{dr}{dN} &= -3r(1-r+m)\left[\frac{1 - w(2+3w)(1-\Omega_m) + (1-\Omega_m)w'}{1-3w(1-\Omega_m)}\right]\,,
\\
\frac{dm}{dN} &= -3m(1-m+m_2)\left[\frac{1 - w(2+3w)(1-\Omega_m) + (1-\Omega_m)w'}{1-3w(1-\Omega_m)}\right]\,,
\\
\frac{dm_2}{dN} &= -3m_2(1-m_2+m_3)\left[\frac{1 - w(2+3w)(1-\Omega_m) + (1-\Omega_m)w'}{1-3w(1-\Omega_m)}\right]\,,
\\
& .................................... \nonumber
\\
& .................................... \nonumber 
\\
\frac{dm_i}{dN} &= -3m_i(1-m_i+m_{i+1})\left[\frac{1 - w(2+3w)(1-\Omega_m) + (1-\Omega_m)w'}{1-3w(1-\Omega_m)}\right]\,.
\end{align}
\end{subequations}

The physically viable domain of the phase space is given by 
\begin{align}\label{phys_viab_f(R)_2}
& {\cal D} = \nonumber\\
& \large\lbrace (\Omega_m,r,m,w,w',.......)\in\mathbb{R}^{i+4}: 0<\Omega_m<1 \,\,\wedge\,\,m\geq0\,\,\wedge\,\, \nonumber \\
& \frac{r \left(6 m (w (3 w+2) (\Omega_m-1)+1)+9 w^2 (\Omega_m-1)^2-1\right)-6 m r (\Omega_m-1) w'-(3 w (\Omega_m-1)+1)^2}{2 r (3 w (\Omega_m-1)+1)}\leq 0 \large\rbrace\,.
\end{align}
with the last inequality coming from the Friedmann constraint \eqref{constr_f(R)_2}.

Notice that the dynamical system \eqref{DS_f(R)_3} is singular in both the matter domination limit $\Omega_m\to1$ and the GR limit $m\to0$. If we confine ourselves within the physically viable region of the phase space where the conditions $0<\Omega_m<1$ and $m>0$ holds, the dynamical system \eqref{DS_f(R)_3} can be regularized to accommodate both the limits, by a time redefinition on the phase space of the form
\begin{equation}
    dN \to d\tilde{N}=\frac{dN}{m(1-\Omega_m)}\,.
\end{equation}
In terms of the new time variable $\tilde{N}$, the dynamical system \eqref{DS_f(R)_3} can be rewritten as
\begin{subequations}\label{DS_f(R)_4}
\begin{align}
\frac{\Omega_m}{f_R} &= -\frac{r \left(6 m (w (3 w+2) (\Omega_m-1)+1)+9 w^2 (\Omega_m-1)^2-1\right)-6 m r (\Omega_m-1) w'-(3 w (\Omega_m-1)+1)^2}{2 r (3 w (\Omega_m-1)+1)}\,,
\\
\frac{d\Omega_m}{d\tilde{N}} &= 3\,w\,m\,\Omega_m(1-\Omega_m)^2\,,
\\
\frac{dw}{d\tilde{N}} &= m\,w'(1-\Omega_m)\,,
\\
\frac{dw'}{d\tilde{N}} &= -\frac{1}{2} \Bigg[(w(\Omega_m - 1) + 1)(3w(\Omega_m - 1) + 1) - \frac{(3w(\Omega_m - 1) + 1)^2}{r} \nonumber \\
&\qquad - \frac{6\,m\,m_2 \left(-(\Omega_m - 1)w' + w(3w + 2)(\Omega_m - 1) + 1\right)^2}{1-3w(1-\Omega_m)} - 5m\Bigg] \nonumber \\
&\qquad - \frac{3}{2}\,m\,w(3w^2 + 2w - w')(1-\Omega_m)^2 - \frac{1}{2}\,m\,(18w^3 + 27w^2 - 18ww' - 9w' + w)(1-\Omega_m)\,,
\\
\frac{dr}{d\tilde{N}} &= -3\,r\,m(1-r+m)(1-\Omega_m)\left[\frac{1 - w(2+3w)(1-\Omega_m) + (1-\Omega_m)w'}{1-3w(1-\Omega_m)}\right]\,,
\\
\frac{dm}{d\tilde{N}} &= -3\,m^2(1-m+m_2)(1-\Omega_m)\left[\frac{1 - w(2+3w)(1-\Omega_m) + (1-\Omega_m)w'}{1-3w(1-\Omega_m)}\right]\,,
\\
\frac{dm_2}{d\tilde{N}} &= -3\,m\,m_2(1-m_2+m_3)(1-\Omega_m)\left[\frac{1 - w(2+3w)(1-\Omega_m) + (1-\Omega_m)w'}{1-3w(1-\Omega_m)}\right]\,,
\\
& .................................... \nonumber
\\
& .................................... \nonumber 
\\
\frac{dm_i}{d\tilde{N}} &= -3\,m\,m_i(1-m_i+m_{i+1})(1-\Omega_m)\left[\frac{1 - w(2+3w)(1-\Omega_m) + (1-\Omega_m)w'}{1-3w(1-\Omega_m)}\right]\,.
\end{align}
\end{subequations}

The Hu-Sawicki example in Sec.~\ref{sec:HS} shows how the $m_i$-hierarchy gives a theory closure even when different branches share the same $r$ in the traditional $m(r)$ formulation \cite{Amendola:2006we,Carloni:2007br,Bahamonde:2017ize,Chakraborty:2021mcf}.

\subsection{Closing for a given cosmology}\label{subsec:cosmography_f(R)}

In this subsection, we present the construction of how a dynamical system can be formulated once one is supplied, not with a given theory $f(R)$, but with a cosmology specified by a cosmographic constraint. This subsection essentially formalizes the closure idea first introduced in \cite{Chakraborty:2021jku} to study the so-called $\Lambda$CDM-mimicking $f(R)$ gravity theories, and later on utilized in \cite{Arora:2022dti,Chakraborty:2023clu} in other contexts. 

We rewrite the dynamical system as
\begin{subequations}\label{DS_f(R)_5}
\begin{align}
\frac{\Omega_m}{f_R} &= 1 + (1-q)\left(\frac{1}{r}-1\right) + m\left(\frac{j-q-2}{1-q}\right)\,,
\\
\frac{dr}{dN} &= r(1-r+m)\left(\frac{j-q-2}{1-q}\right)\,,
\\
\frac{dm}{dN} &= \frac{ - m(1-q)r\,s - 3(1-q)^3 + r\left[m \left(j^2 - (j-2)^2\,m - 6j + 2\right) + 2\right]}{(1-q)(j-q-2)r}\,\nonumber 
\\
&\quad + \frac{q\,r \left[m^2 (2j-q-4) + m\,q(q+6) - (q-4)q - 5\right]}{(1-q)(j-q-2)r}
\\
\frac{dq}{dN} &= q(1+2q) - j
\\
\frac{dj}{dN} &= j(2+3q) + s\,,
\\
& .................................... \nonumber
\\
& .................................... \nonumber 
\\
\frac{dc_i}{dN} &= [(i+2)(1-c_0)-1]c_i + c_{i+1}\,,
\end{align}
\end{subequations}

The physically viable domain of the phase space is given by
\begin{equation}
    {\cal D} = \left\lbrace \{r,m,q,j,....,c_i\}\in \mathbb{R}^{i+3}: m\geq0\,\,\wedge\,\,1 + (1-q)\left(\frac{1}{r}-1\right) + m\left(\frac{j-q-2}{1-q}\right)>0 \right\rbrace\,.
\end{equation}

It was argued in \cite{Chakraborty:2023clu} that if a given cosmic solution $a(t)$ is believed to be the solution of some $f(R)$ gravity with the assumption of spatial flatness and in the presence of a barotropic perfect fluid with a constant equation of state parameter, then one can express the cosmographic snap parameter $c_2\equiv s$ algebraically in terms of the lower order cosmographic parameters $\{c_0,c_1\}\equiv\{-q,j\}$; see the discussion in \cite[Sec.~3]{Chakraborty:2023clu}. In that sense, the above hierarchy goes maximum up to the $dj/dN$-equation.

Many of the physically interesting cosmological evolutions, however can already be expressed in the form $j=j(q)$. When $j$ can be expressed as a function of $q$, the cosmographic hierarchy truncates at the level of the $dq/dN$ equation itself. All the higher-order cosmographic parameters can be expressed as a function of $q$. 

A trivial example is the $\Lambda$CDM-like cosmic evolution, which is given by $j=1$ \cite{Chakraborty:2021jku}. An \emph{almost} $\Lambda$CDM-like cosmic evolution can typically be expressed in the form $j=j(q,\epsilon)$, where the parameter $\epsilon$ quantifying the deviation from an exact $\Lambda$CDM-like evolution can be constrained using a combination of data sets \cite{Worsley:2026ijo}. 

On the other hand, in a late-time cosmology, $q$ is a monotonically decreasing quantity with time, so that ``$-q$'' can be considered as a proxy time variable. In such cases, it is sometimes more illuminating to write the above system in the form
\begin{subequations}\label{DS_f(R)_j(q)}
\begin{align}
\frac{dr}{d(-q)} &= r(1-r+m)\left[\frac{j(q)-q-2}{(1-q)[j(q)-q(1+2q)]}\right]\,,
\\
\frac{dm}{d(-q)} &= \frac{ - m(1-q)r\,s(q) - 3(1-q)^3 + r\left[m \left(j^2(q) - (j(q)-2)^2\,m - 6j(q) + 2\right) + 2\right]}{r(1-q)(j(q)-q-2)[j(q)-q(1+2q)]}\,\nonumber 
\\
&\quad + \frac{q\,r \left[m^2 (2j(q)-q-4) + m\,q(q+6) - (q-4)q - 5\right]}{r(1-q)(j(q)-q-2)[j(q)-q(1+2q)]}\,.
\end{align}
\end{subequations}
The non-autonomous system above dictates the flow of the solution curves in the theory space $r-m$ for $f(R)$ theories that reproduces a given cosmology $j=j(q)$. This is the approach taken recently in \cite{Chakraborty:2025lkz}. 

\subsection{Closing for a given dark energy E.o.S. parametrization}\label{subsec:eos_fR}

This subsection is dedicated to presenting a dynamical system formulation such that one can embed a given dark energy equation of state parametrization within the $f(R)$ framework. The idea is to study those $f(R)$ theories which reproduces a chosen form of dark energy equation of state evolution. Embedding a dark energy equation of state parametrization within the autonomous system framework has previously been explored in the context of DBI-tachyonic dark energy models in \cite{Hussain:2022dhp}. 

We rewrite the dynamical system as
\begin{subequations}\label{DS_f(R)_6}
\begin{align}
\frac{\Omega_m}{f_R} &= -\frac{r \left(6 m (w (3 w+2) (\Omega_m-1)+1)+9 w^2 (\Omega_m-1)^2-1\right)-6 m r (\Omega_m-1) w'-(3 w (\Omega_m-1)+1)^2}{2 r (3 w (\Omega_m-1)+1)}\,,
\\
\frac{d\Omega_m}{dN} &= 3w\Omega_m(1-\Omega_m)\,,
\\
\frac{dr}{dN} &= - 3r(1-r+m)\left[\frac{1 - w(2+3w)(1-\Omega_m) + (1-\Omega_m)w'}{1-3w(1-\Omega_m)}\right]\,,
\\
\frac{dm}{dN} &= \frac{-r m\,P + 6 r m^2 B^2 + A^2\bigl[(r-3)w(1-\Omega_m)-r+1\bigr]}{2 r A B}\,,
\\
\frac{dw}{dN} &= w' \,,
\\
\frac{dw'}{dN} &= w''\,,
\\
& .................................... \nonumber
\\
& .................................... \nonumber 
\end{align}
\end{subequations}
where
\begin{equation}
    A = 3 w (\Omega_m - 1) + 1,
\qquad
B = -(\Omega_m - 1) w' + w (3 w + 2)(\Omega_m - 1) + 1,
\end{equation}
and
\begin{equation}
    \begin{aligned}
P ={}& 27 \Omega_m^3 w^4 + 18 \Omega_m^3 w^3 - 9 \Omega_m^3 w^2 w'
- 81 \Omega_m^2 w^4 - 54 \Omega_m^2 w^3 + 45 \Omega_m^2 w^2 w' \\
&+ 27 \Omega_m^2 w^2 - 6 \Omega_m^2 w w'' + 6 \Omega_m^2 (w')^2
+ 81 \Omega_m w^4 + 36 \Omega_m w^3 - 63 \Omega_m w^2 w' \\
&- 45 \Omega_m w^2 + 18 \Omega_m w w' + 12 \Omega_m w w'' + 8 \Omega_m w
- 12 \Omega_m (w')^2 - 3 \Omega_m w' \\
&- 2 \Omega_m w'' - 27 w^4 + 27 w^2 w' + 18 w^2
- 18 w w' - 6 w w'' - 8 w \\
&+ 6 (w')^2 + 3 w' + 2 w'' + 1.
\end{aligned}
\end{equation}

The physically viable domain of the phase space is given by 
\begin{align}
& {\cal D} = \nonumber\\
& \large\lbrace (\Omega_m,r,m,w,w',.......): 0<\Omega_m<1 \,\,\wedge\,\,m\geq0\,\,\wedge\,\, \nonumber \\
& \frac{r \left(6 m (w (3 w+2) (\Omega_m-1)+1)+9 w^2 (\Omega_m-1)^2-1\right)-6 m r (\Omega_m-1) w'-(3 w (\Omega_m-1)+1)^2}{2 r (3 w (\Omega_m-1)+1)}\leq 0 \large\rbrace\,.
\end{align}

For many of the physically interesting dark energy equation of state parametrizations, the hierarchy $\{w,w',w''....\}$ can be truncated at the level of $w'$ itself. A prime example is the CPL parametrization $w=w_0+w_a(1-a)$, which can be expressed as
\begin{equation}\label{CPL}
    w'(w)=w-w_{(z\to\infty)}\,.
\end{equation}

On the other hand, in a late-time cosmology, $\Omega_m$ is a monotonically decreasing quantity with time, so that ``$-\Omega_m$'' can be considered as a proxy time variable. In such cases, it is sometimes more illuminating to write the above system in the form
\begin{subequations}
\begin{align}
\frac{dr}{d(-\Omega_m)} &= r(1-r+m)\left(\frac{1 - w(2+3w)(1-\Omega_m) + (1-\Omega_m)w'(w)}{w\Omega_m(1-\Omega_m)[1-3w(1-\Omega_m)]}\right)\,,
\\
\frac{dm}{d(-\Omega_m)} &= \frac{r m\,P - 6 r m^2 B^2 - A^2\bigl[(r-3)w(1-\Omega_m)-r+1\bigr]}{2 r A B [3w\Omega_m(1-\Omega_m)]}\,,
\\
\frac{dw}{d(-\Omega_m)} &= -\frac{w'(w)}{3w\Omega_m(1-\Omega_m)} \,.
\end{align}
\end{subequations}

For a constant dark energy equation of state $w$; $w=w_c$, the autonomous system can be simplified even further:
\begin{subequations}\label{DS_f(R)_wc}
\begin{align}
\frac{dr}{d(-\Omega_m)} &= r(1-r+m)\left(\frac{1 - w_c(2+3w_c)(1-\Omega_m)}{w_c\Omega_m(1-\Omega_m)[1-3w_c(1-\Omega_m)]}\right)\,,
\\
\frac{dm}{d(-\Omega_m)} &= \frac{r m\,P(w_c,\Omega_m) - 6 r m^2 B^2(w_c,\Omega_m) - A^2(w_c,\Omega_m)\bigl[(r-3)w_c(1-\Omega_m)-r+1\bigr]}{2 r A(w_c,\Omega_m) B(w_c,\Omega_m) [3w_c\Omega_m(1-\Omega_m)]}\,.
\end{align}
\end{subequations}
where
\begin{equation}
    A(w_c,\Omega_m) = 3 w_c (\Omega_m - 1) + 1,
\qquad
B(w_c,\Omega_m) = w_c (3 w_c + 2)(\Omega_m - 1) + 1,
\end{equation}
and
\begin{equation}
    \begin{aligned}
P ={}& 27 \Omega_m^3 w_c^4 + 18 \Omega_m^3 w_c^3 - 81 \Omega_m^2 w_c^4 - 54 \Omega_m^2 w_c^3 \\
&+ 27 \Omega_m^2 w_c^2 + 81 \Omega_m w_c^4 + 36 \Omega_m w_c^3 - 45 \Omega_m w_c^2 + 8 \Omega_m w_c - 27 w_c^4 + 18 w_c^2 - 8 w_c + 1.
\end{aligned}
\end{equation}
The non-autonomous system above, therefore, dictates the flow of the solution curves in the theory space $r-m$ for those $f(R)$ theories that mimic a constant equation of state dark energy component.

\section{$f(Q)$ gravity}\label{sec:f(Q)}

For the trivial FLRW connection branch, which is many a times loosely termed as the so-called coincident gauge for $f(Q)$ cosmology, one has $Q=-6H^2$. The field equations can be written in a manner parallel to $f(R)$:
\begin{subequations}\label{fieldeqs_f(Q)}
    \begin{align}
        3H^{2} &= \kappa\rho_{\rm tot} = \kappa\rho + \kappa\rho_{\rm DE}\,,
        \\
        -\left(2\dot{H} + 3H^{2}\right) &= \kappa P_{\rm tot} = \kappa P_{\rm DE}\,,
    \end{align}     
\end{subequations}
with
\begin{subequations}\label{nonmet_fluid}
    \begin{align}
        \kappa\rho_{\rm DE} &= \frac{1}{2}(Qf_Q - f) + 3H^2(1-f_Q)\,,\label{DE_ed_f(Q)}
        \\
        \kappa P_{\rm DE} &= 2H\dot f_Q - \frac{1}{2}(Qf_Q-f) - (2\dot{H}+3H^2)(1-f_Q)\,.\label{DE_p_f(Q)}
    \end{align}
\end{subequations}
Notice that, unlike the $f(R)$ case, there is no $\ddot{f}_Q$ term here: $f(Q)$ gravity in the coincident gauge is second order, so only $\dot f_Q$ ever appears. Since this is again just a matter-plus-effective-fluid split with pressureless matter, the dark energy equation of state and the conservation equations are exactly similar as before:
\begin{equation}\label{DE_eos_f(Q)}
    w = \frac{P_{\rm DE}}{\rho_{\rm DE}} = \frac{2H\dot f_Q - \frac{1}{2}(Qf_Q-f) - (2\dot{H}+3H^2)(1-f_Q)}{\frac{1}{2}(Qf_Q - f) + 3H^2(1-f_Q)} = \frac{2}{3}\left(\frac{q-1/2}{1-\Omega_m}\right)\,,
\end{equation}
and
\begin{subequations}\label{cons_f(Q)}
\begin{eqnarray}
    && \dot\rho+3H\rho=0\,,
    \\
    && \dot\rho_{\rm DE}+3H\rho_{\rm DE}(1+w)=0\,.
\end{eqnarray}
\end{subequations}

\subsection{Closing by a given theory}\label{subsec:theory_f(Q)}

With $m_0\equiv r = Qf_Q/f$ and $m_1\equiv m = Qf_{QQ}/f_Q$, the Friedmann equation gives \cite{Dutta:2025fqw}
\begin{equation}\label{fried-f(Q)}
    \frac{\Omega_m}{f_Q} = 2 - \frac{1}{r}\,,
\end{equation}
while the Raychaudhuri equation gives \cite{Dutta:2025fqw}
\begin{equation}\label{raych-f(Q)}
    m = -\frac{1}{2} + \frac{3}{2r}\left(\frac{r-1/2}{1+q}\right)\,,
    \qquad\text{equivalently}\qquad
    q = \frac{4r(1-m)-3}{2r(1+2m)}\,.
\end{equation}
Because $Q=-6H^2$ depends on $H$ alone (not on $\dot H$), $\dot Q/(QH)=-2(1+q)$ needs only $q$ -- and $q$ is already fixed algebraically by $r,m$ through \eqref{eq:fQ_m_of_rq}. This is the key structural simplification relative to $f(R)$. Since the closure coefficient never needed $j$ to begin with, the entire theory-space hierarchy closes on $\{r,m,m_2,\ldots\}$ by itself:
\begin{subequations}\label{DS_f(Q)_1}
\begin{align}
\frac{\Omega_m}{f_Q} &= 2 - \frac{1}{r}\,,
\\
\frac{dr}{dN} &= 3(1-r+m)\left(\frac{1-2r}{1+2m}\right)\,,
\\
\frac{dm}{dN} &= 3(1-m+m_2)\frac{m}{r}\left(\frac{1-2r}{1+2m}\right)\,,
\\
\frac{dm_2}{dN} &= 3(1-m_2+m_3)\frac{m_2}{r}\left(\frac{1-2r}{1+2m}\right)\,,
\\
& .................................... \nonumber
\\
& .................................... \nonumber
\\
\frac{dm_i}{dN} &= 3(1-m_i+m_{i+1})\frac{m_i}{r}\left(\frac{1-2r}{1+2m}\right)\,.
\end{align}
\end{subequations}

Although the condition $f_Q>0$ can be demanded from the ground that the effective gravitational coupling has to be positive, there is no such physical ground to demand the condition $f_{QQ}>0$ for physical viability; the condition for tachyonic stability cannot be reduced to such a simple condition for $f(Q)$ gravity. The physically viable domain of the phase space is therefore given by
\begin{equation}\label{phys_viab_f(Q)_1}
{\cal D} = \{(r,m,m_2,....,m_i)\in\mathbb{R}^{i+1}:r<0\,\,\vee\,\,r\geq1/2\}\,,
\end{equation}
coming simply from the Friedmann constraint \eqref{fried-f(Q)}.

Notice that $q,\,j,\,w,\,w'$ etc do not enter the system dynamically anymore, but rather are already determined algebraically with respect to the $m_i$-parameters:
\begin{subequations}
\begin{align}
q &= \frac{4r(1-m)-3}{2r(1+2m)}\,,
\label{eq:fQ_m_of_rq}\\
j &= \frac{2r^2(2m+1)^3 - 9m(2r-1)^2 (3+2m_2)}{2r^2(2m+1)^3}\,,
\label{eq:fQ_j_of_rmm2}\\
w &= -\frac{1 + (2m-1)r}{(1+2m)(1-\Omega_m)r}\,,
\\
w' &=
\frac{3}{(1+2m)^3(1-\Omega_m)^2r^2}
\Bigg[
8m^3\Omega_m r^2
-(r-1)\left(1+(\Omega_m-2)r\right)
+4m^2r\left(1+\Omega_m+(\Omega_m-2)r\right)
\nonumber\\
&\qquad\qquad\qquad
+m\left(
5-3\Omega_m
+2m_2(1-\Omega_m)(1-2r)^2
-16(1-\Omega_m)r
+2(6-7\Omega_m)r^2
\right)
\Bigg].
\end{align}
\end{subequations}
Expressions for the higher-order cosmographic snap parameter $s$, or that of $w'''$, will contain the theory-space parameter $m_3$.

\subsection{Closing for a given cosmology}\label{subsec:cosmography_f(Q)}

The system above keeps $\{r,m,m_2,\ldots\}$ as the fundamental phase-space variables, with $q$ and $j$ read off algebraically from them via \eqref{eq:fQ_m_of_rq} and \eqref{eq:fQ_j_of_rmm2}. One can instead run the substitution the other way: solving the Raychaudhuri relation \eqref{raych-f(Q)} for $r$ rather than $q$ gives
\begin{equation}\label{eq:fQ_r_of_qm}
    r = \frac{3}{2\left[2(1-m)-q(1+2m)\right]}\,,
\end{equation}
which trades $r$ for $q$ at fixed $m$ (one checks easily that substituting \eqref{eq:fQ_r_of_qm} back into \eqref{eq:fQ_m_of_rq} returns $q$ identically). Since $q$ is now available as a genuine phase-space variable, it is natural to also invert \eqref{eq:fQ_j_of_rmm2} for $m_2$ and substitute $q\to q(r,m)$; this gives
\begin{equation}\label{eq:fQ_m2_companion}
    m_2 = -\frac{3}{2} - \frac{j-1}{(1+q)^2}\left(\frac{1+2m}{4m}\right)\,.
\end{equation}

Substituting \eqref{eq:fQ_r_of_qm} into the Friedmann constraint \eqref{fried-f(Q)}, and both \eqref{eq:fQ_r_of_qm} and \eqref{eq:fQ_m2_companion} into the $dm/dN$ equation from the system \eqref{DS_f(Q)_1}, we get after simplification
\begin{equation}\label{eq:fQ_dmdN_qjm}
    \frac{\Omega_m}{f_Q} = \frac{2}{3}(1+2m)(1+q)\,,
    \qquad
    \frac{dm}{dN} = \frac{(1+2m)\left[(j-1)+2m(1+q)^2\right]}{2(1+q)}\,.
\end{equation}
Together with the cosmographic recursion relation \eqref{cosm_recursion}, which is completely kinematic in nature and is oblivious to the underlying theory, the closed system reads
\begin{subequations}\label{DS_f(Q)_2}
\begin{align}
\frac{\Omega_m}{f_Q} &= \frac{2}{3}(1+2m)(1+q)\,,
\\
\frac{dm}{dN} &= \frac{(1+2m)\left[(j-1)+2m(1+q)^2\right]}{2(1+q)}\,,
\\
\frac{dq}{dN} &= q(1+2q) - j\,,
\\
\frac{dj}{dN} &= j(2+3q) + s\,,
\\
& .................................... \nonumber
\\
& .................................... \nonumber
\\
\frac{dc_i}{dN} &= [(i+2)(1-c_0)-1]c_i + c_{i+1}\,.
\end{align}
\end{subequations}
This is now structurally parallel to the $f(R)$ system of the previous subsection, with the $dr/dN$ equation simply absent. $r$ is no longer needed as an independent phase-space variable at all, having been traded for $q$.

The physically viable domain of the phase space is given by
\begin{equation}\label{phys_viab_f(Q)_2}
{\cal D} = \{(m,q,j,....,c_i)\in\mathbb{R}^{i+2}:(1+2m)(1+q)>0\}\,.
\end{equation}

As in the $f(R)$ case, fixing the cosmographic hierarchy at successively higher order closes this system at successively higher, but finite, dimension. In particular, if a full ansatz $j=j(q)$ is given, $dj/dN$ is no longer needed and the system closes on $\{q,m\}$: two dimensions.

\subsection{Closing for a given dark energy E.o.S. parametrization}\label{subsec:eos_fQ}

Let us start from the system \eqref{DS_f(Q)_2}, trade $q,\,j$ in terms of $\Omega_m,\,w,\,w'$ using the statefinder relations \eqref{q}, \eqref{j}. Then it is possible to rewrite the dynamical system \eqref{DS_f(Q)_2} as
\begin{subequations}\label{DS_f(Q)_3}
\begin{align}
\frac{\Omega_m}{f_Q} &= (1+2m)(1 + w(1-\Omega_m))\,,
\\
\frac{d\Omega_m}{dN} &= 3w\Omega_m(1-\Omega_m)\,,
\\
\frac{dm}{dN} &= \frac{(1+2m)\left[3m(1 + w(1-\Omega_m))^2 + (1-\Omega_m)(3w^2+3w-w')\right]}{2(1 + w(1-\Omega_m))}\,,
\\
\frac{dw}{dN} &= w'\,,
\\
\frac{dw'}{dN} &= w''\,,
\\
& .................................... \nonumber
\end{align}
\end{subequations}
again structurally parallel to the $f(R)$ system with $dr/dN$ absent. Notice however that unlike the $f(R)$ case, the expression for $dm/dN$ needs only $w'$, not $w''$. So here the $\{w,w',w'',\ldots\}$ hierarchy opens up one order more slowly than for $f(R)$, where the analogous $dm/dN$ equation needed $w''$ from the start.

The physically viable domain of the phase space is given by
\begin{equation}\label{phys_viab_f(Q)_3}
{\cal D} = \{(\Omega_m,m,w,w',....):(1+2m)(1 + w(1-\Omega_m))>0\}\,.
\end{equation}

As for $f(R)$, this becomes a genuine truncation only once a specific $w'=w'(w)$ ansatz is imposed, e.g., the CPL-motivated relation $w'(w)=w-w_{(z\to\infty)}$ \eqref{CPL}. Most simply, a constant equation of state $w=w_c$ (so that $w'=0$), the system closes on $\{\Omega_m,m\}$:
\begin{subequations}\label{DS_f(Q)_wc}
\begin{align}
\frac{d\Omega_m}{dN} &= 3w_c\Omega_m(1-\Omega_m)\,,
\\
\frac{dm}{dN} &= \frac{3(1+2m)\left[m(1 + w_c(1-\Omega_m))^2+w_c(1+w_c)(1-\Omega_m)\right]}{2(1 + w_c(1-\Omega_m))}\,.
\end{align}
\end{subequations}

\section{Application for a given $f(R)$ model: Hu-Sawicki}\label{sec:HS}

As an application of the dynamical system formulation \eqref{DS_f(R)_1} and \eqref{DS_f(R)_3}, which provides two alternative dynamical system formulation once a theory function $f(R)$ is supplied, let us turn towards a very well-known late time $f(R)$ model -- the Hu-Sawicki model. The generic Hu-Sawicki $f(R)$ model is
\begin{equation}
    f(R) = R - \frac{C_1 R^n}{1+C_2 R^n}\,.
\end{equation}
One can verify that, for the case $n=1=C_1$, one has
\begin{equation}
    m_0(r)\equiv r(R) = \frac{C_2 R +2}{C_2 R +1}\,, \qquad m_1(r)\equiv m(R) = \frac{2}{(C_2 R +1)(C_2 R +2)}\,.
\end{equation}
One can see that the function $r=r(R)$ is uniquely invertible to obtain $R=R(r)$, substituting which into the expression of $m(R)$, one gets a unique relation
\begin{equation}
    m(r) = \frac{2(r-1)^2}{r}\,.
\end{equation}
A comprehensive dynamical system analysis of the Hu-Sawicki case with $n=1=C_1$ has been performed in \cite{MacDevette:2022hts}

To establish the merit of the dynamical system framework that we present here, utilizing the $m_i$-hierarchy, over the standard dynamical system framework that is utilized in, say, \cite{MacDevette:2022hts}, let us actually implement it somewhere where the standard dynamical system formulation is known to fail. Therefore, we decide to keep all three parameters of the Hu-Sawicki model completely free -- except the only restrictions that they are all positive $n,C_1,C_2>0$ -- and investigate its cosmological phase space in full generality. One gets
\begin{align}
    m_0(R) \equiv r(R)
    &= \frac{(1+C_2R^n)^2-nC_1R^{n-1}}
            {(1+C_2R^n)(1+C_2R^n-C_1R^{n-1})} \,, \\[6pt]
    m_1(R) \equiv m(R)
    &= \frac{nC_1R^{n-1}\bigl[(n+1)C_2R^n-(n-1)\bigr]}
            {(1+C_2R^n)\bigl[(1+C_2R^n)^2-nC_1R^{n-1}\bigr]} \,, \\[6pt]
    m_2(R)
    &= \frac{(n-1)(n-2)-4(n^2-1)C_2R^n+(n+1)(n+2)C_2^2R^{2n}}
            {(1+C_2R^n)\bigl[(n-1)-(n+1)C_2R^n\bigr]} \,, \\[6pt]
    m_3(R)
    &= \frac{\begin{aligned}
          &(n-1)(n-2)(n-3)-(n^2-1)(11n-18)C_2R^n \\
          &\quad +(n^2-1)(11n+18)C_2^2R^{2n}
                 -(n+1)(n+2)(n+3)C_2^3R^{3n}
        \end{aligned}}
        {(1+C_2R^n)\bigl[(n-1)(n-2)-4(n^2-1)C_2R^n
                         +(n+1)(n+2)C_2^2R^{2n}\bigr]} \,.
\end{align}
In the high curvature limit, one has
\begin{align}
r(R)
&= 1+\frac{C_1}{C_2 R}+o(R^{-1})\,,\\
m(R)
&= \frac{n(n+1)C_1}{C_2^2 R^{n+1}}
   +o\!\left(R^{-(n+1)}\right)\,,\\
m_2(R)
&= -(n+2)+O\!\left(R^{-n}\right)\,,\\
m_3(R)
&= -(n+3)+O\!\left(R^{-n}\right)\,,
\end{align}
from which one can see that $m\to0$, i.e., the theory generically approaches GR in the high curvature limit.

Generically, however, one can now see that the expression $r=r(R)$ is not uniquely invertible to obtain $R=R(r)$. Consequently, one cannot obtain a unique function $m=m(r)$ here, which apparently leaves the dynamical system analysis in jeopardy. However, with some rigorous mathematical analysis that we leave for the appendix \ref{app:HS_constraint_surface}, one can actually eliminate all three free parameters $n,\,C_1,\,C_2$, and arrive at an algebraic relation of the form
\begin{align}
\mathcal{F}(r,m,m_2,m_3)
={}&
\left[
\begin{aligned}
    &9r^2m^2(2m_2-9) \\
    &+6rm(r-1)\bigl[27+m_2(4m_2-6m_3+3)\bigr] \\
    &+(r-1)^2m_2\bigl[9(2m_2+1)(m_3+1)-2m_2(8m_2+3)\bigr]
\end{aligned}
\right]^2
\nonumber\\
&-\bigl(16m_2^2-12m_2m_3+9\bigr)
\left[
\begin{aligned}
    &-9r^2m^2+6rm(r-1)(m_2+3) \\
    &+(r-1)^2m_2(3m_3+3-4m_2)
\end{aligned}
\right]^2
=0 \,.\label{eq:HS_constraint}
\end{align}
The above algebraic equation represents the Hu-Sawicki constraint surface in the theory space. Any given Hu-Sawicki theory, for any values of the free model parameters $n,\,C_1,\,C_2$, must reside on this surface within the theory space spanned by the parameters $\{m_i\}$. For consistency, this constraint must be preserved in time; $d{\cal F}/dN=0$, which then provides a differential equation for $m_3$, which allows us to close the system as written below.

Following the system \eqref{DS_f(R)_2}, the full closed dynamical system for the generic Hu-Sawicki model can be expressed as
\begin{subequations}\label{DS_HS_1}
\begin{align}
\frac{\Omega_m}{f_R} &= 1 + (1-q)\left(\frac{1}{r}-1\right) + m\left(\frac{j-q-2}{1-q}\right)\,,
\\
\frac{dq}{d\tilde{N}} &= m(2q^2 + q - j)\,,
\\
\frac{dj}{d\tilde{N}} &= -m[2(1-q^2)-3q(j-q-2)] + (1-q)(2-q) - \frac{3(1-q)^2}{r} - \frac{(j-q-2)^2}{(1-q)}\,m\,m_2
\\
\frac{dr}{d\tilde{N}} &= r\,m(1-r+m)\left(\frac{j-q-2}{1-q}\right)\,,
\\
\frac{dm}{d\tilde{N}} &= m^2(1-m+m_2)\left(\frac{j-q-2}{1-q}\right)\,,
\\
\frac{dm_2}{d\tilde{N}} &= m\,m_2\left(1-m_2+m_3\right)\left(\frac{j-q-2}{1-q}\right)\,,
\\
\frac{dm_3}{dN} &=
m\Biggl\{(m_3+2)\left(\frac{5}{3}m_2-m_3\right)
\nonumber\\
&\quad + \left(2m_3+1-\frac{8}{3}m_2\right)
\frac{m_2(r-1)(m_3+2)\bigl[m_2(r-1)-6rm\bigr] +9rm\bigl[rm(m_2-1)+2(r-1)\bigr]    
}{m_2(r-1)^2(4m_2-3m_3-3) -6rm(r-1)(m_2+3)+9r^2m^2
}
\Biggr\}\left(\frac{j-q-2}{1-q}\right)\,.
\end{align}
\end{subequations}
Or, alternatively, following the system \eqref{DS_f(R)_4}, as
\begin{subequations}\label{DS_HS_2}
\begin{align}
\frac{\Omega_m}{f_R} &= -\frac{r \left(6 m (w (3 w+2) (\Omega_m-1)+1)+9 w^2 (\Omega_m-1)^2-1\right)-6 m r (\Omega_m-1) w'-(3 w (\Omega_m-1)+1)^2}{2 r (3 w (\Omega_m-1)+1)}\,,
\\
\frac{d\Omega_m}{d\tilde{N}} &= 3\,w\,m\,\Omega_m(1-\Omega_m)^2\,,
\\
\frac{dw}{d\tilde{N}} &= m\,w'(1-\Omega_m)\,,
\\
\frac{dw'}{d\tilde{N}} &= -\frac{1}{2} \Bigg[(w(\Omega_m - 1) + 1)(3w(\Omega_m - 1) + 1) - \frac{(3w(\Omega_m - 1) + 1)^2}{r} \nonumber \\
&\qquad - \frac{6\,m\,m_2 \left(-(\Omega_m - 1)w' + w(3w + 2)(\Omega_m - 1) + 1\right)^2}{1-3w(1-\Omega_m)} - 5m\Bigg] \nonumber \\
&\qquad - \frac{3}{2}\,m\,w(3w^2 + 2w - w')(1-\Omega_m)^2 - \frac{1}{2}\,m\,(18w^3 + 27w^2 - 18ww' - 9w' + w)(1-\Omega_m)\,,
\\
\frac{dr}{d\tilde{N}} &= -3\,r\,m(1-r+m)(1-\Omega_m)\left[\frac{1+(1-\Omega_m)(w'-2w-3w^2)}
          {1-3w(1-\Omega_m)}\right]\,,
\\
\frac{dm}{d\tilde{N}} &= -3\,m^2(1-m+m_2)(1-\Omega_m)\left[\frac{1+(1-\Omega_m)(w'-2w-3w^2)}{1-3w(1-\Omega_m)}\right]\,,
\\
\frac{dm_2}{d\tilde{N}} &= -3\,m\,m_2\left(1-m_2+m_3\right)(1-\Omega_m)\left[\frac{1+(1-\Omega_m)(w'-2w-3w^2)}
          {1-3w(1-\Omega_m)}\right]\,,
\\
\frac{dm_3}{d\tilde{N}} &=
-m\Biggl\{
    (m_3+2)(5m_2-3m_3)\nonumber\\
    & \quad +(6m_3+3-8m_2)
    \frac{m_2(r-1)(m_3+2)\bigl[m_2(r-1)-6rm\bigr]
          +9rm\bigl[rm(m_2-1)+2(r-1)\bigr]}
         {9r^2m^2+(r-1)\bigl[(r-1)m_2(4m_2-3m_3-3)-6rm(m_2+3)\bigr]}
    \Biggr\}\nonumber\\
    & \quad\times (1-\Omega_m)
     \left[\frac{1+(1-\Omega_m)(w'-2w-3w^2)}
          {1-3w(1-\Omega_m)}\right] \,.
\end{align}
\end{subequations}

An important comment is in order here. In order to avoid any branching structure in the phase space and come up with a unique dynamical system, we decided to take the derivative of the constraint \eqref{eq:HS_constraint} and come up with a $m_3'$-equation. The price to pay is that the dynamical variables $r,\,m,\,m_2,\,m_3$ are not all independent; they are related by the algebraic constraint \eqref{eq:HS_constraint}. A cosmological solution to the generic Hu-Sawicki model must reside on the constraint hypersurface given by \eqref{app:HS_constraint_surface}. The constraint surface provides an additional phase space constraint on top of the standard Friedmann constraint; however, unlike the Friedmann constraint, it cannot be used to conveniently eliminate one variable from the dynamical system without introducing a branching structure in the phase space.

A given phase space point $\{q,j,r,m,m_2,m_3\}$ or $\{\Omega_m,w,w',r,m,m_2,m_3\}$ uniquely determines the theory parameter $n$, but determining $C_1,C_2$ (more precisely, their respective dimensionless versions normalized by $H_0$) uniquely requires one to further use $E\equiv H/H_0$ as a dynamical variable; see appendix \ref{app:HS_parameter_determination}.

If we argue that different Hu-Sawicki families are characterized by different values of the Hu-Sawicki parameter $n$, then we can call the fixed points that exist for any given parameter value $n$ the generic fixed points of Hu-Sawicki. The generic physically viable fixed points (following the condition \eqref{phys_viab_f(R)_1} as well as the additional constraint \eqref{app:HS_constraint_surface}) calculated from the system \eqref{DS_HS_1} are listed in Table \ref{tab:HS_fp_1}.
\begin{table}[H]
\centering
\resizebox{\textwidth}{!}{%
\begin{tabular}{llcl}
\hline\hline
Fixed Point & $(q,j,r,m,m_2,m_3)$
            & Stability & Cosmology \\
\hline
$M_{\rm matter}$
& $\left(1/2,1,1,0,-n-2,-n-3\right)$
& Non-hyperbolic
& GR-like matter domination \\[3pt]
$L_c$
& $(-1,1,2,m_*,m_{2*},m_{3*})$
& \makecell{Sink for $0<m_*<1$\\
            Saddle for $m_*>1$\\
            Non-hyperbolic for $m_*=1$}
& de Sitter \\
\hline\hline
\end{tabular}%
}
\caption{The GR-matter-dominated fixed point at high curvature appearing as a line of non-hyperbolic fixed points and the late-time de-Sitter fixed points appearing as a 3-parameter family of fixed points, as obtained from \eqref{DS_HS_1}. Since $n>0$ in the original Hu-Sawicki model, the matter line exists for $m_2<-2,\,m_3<-3$. The starred coordinates lie on the admissible branch of $\mathcal F(2,m_*,m_{2*},m_{3*})=0$.}
\label{tab:HS_fp_1}
\end{table}

\begin{table}[H]
\centering
\resizebox{\textwidth}{!}{%
\begin{tabular}{llcl}
\hline\hline
Fixed Point & $(\Omega_m,w,w',r,m,m_2,m_3)$
            & Stability & Cosmology ($q$) \\
\hline
$M_{\rm matter}$
& $(1,\text{--},\text{--},1,0,-n-2,-n-3)$
& Non-hyperbolic
& GR-like matter domination $\left(\frac12\right)$ \\[3pt]
$L_c$
& $(0,-1,0,2,m_*,m_{2*},m_{3*})$
& \makecell{Sink for $0<m_*<1$\\
            Saddle for $m_*>1$\\
            Non-hyperbolic for $m_*=1$}
& de Sitter $(-1)$ \\
\hline\hline
\end{tabular}%
}
\caption{The same physical states in the equation-of-state system
\eqref{DS_HS_2}. The dashes in $M_{\rm matter}$ indicate that the statefinder relations do not determine $w,w'$ exactly at the matter dominated fixed point; since the dark energy density vanishes, the dark energy equation of state loses its meaning. }
\label{tab:HS_fp_2}
\end{table}

Both the tables \ref{tab:HS_fp_1} and \ref{tab:HS_fp_2}, which basically show two slightly different ways of representing the fixed points, reveal the existence of a family of fixed points representing the General Relativistic matter dominated epoch $M_{matter}$, and a family of fixed points representing the curvature (dark energy) dominated de-Sitter epoch $L_c$. From the expression of $n$ provided in Eq.\eqref{eq:dynvars_to_HSparams}, one can calculate that $n=-m_{2*}-2=-m_{3*}-3$ for $M_{\rm matter}$ and $n=\frac{4m_*}{2m_*-h_*}-1-h_*$ for $L_c$, with the subscript $``*''$ meaning that the variables are evaluated at the fixed point. This establishes the generic existence of both these fixed points across the entire Hu-Sawicki family. Different values of $n$ correspond to different points belonging to the family $M_{\rm matter}$ or $L_c$.

For a particular Hu-Sawicki family with a specific value of $n$, one could find additional fixed points, which are not investigated here. The $n=1$ case is investigated in detail in \cite{MacDevette:2022hts}.

\begin{figure}[H]
    \centering
    \begin{subfigure}[b]{0.49\linewidth}
    \includegraphics[width=\linewidth]{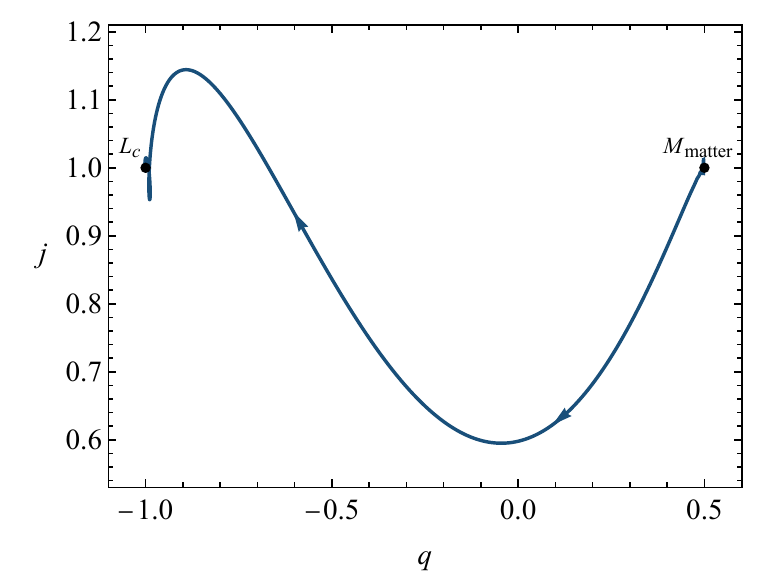}
     \caption{}
     \label{fig:q-j trajectory}
    \end{subfigure}
    \hspace{1.4mm}
    \begin{subfigure}[b]{0.49\linewidth}
    \includegraphics[width=\linewidth]{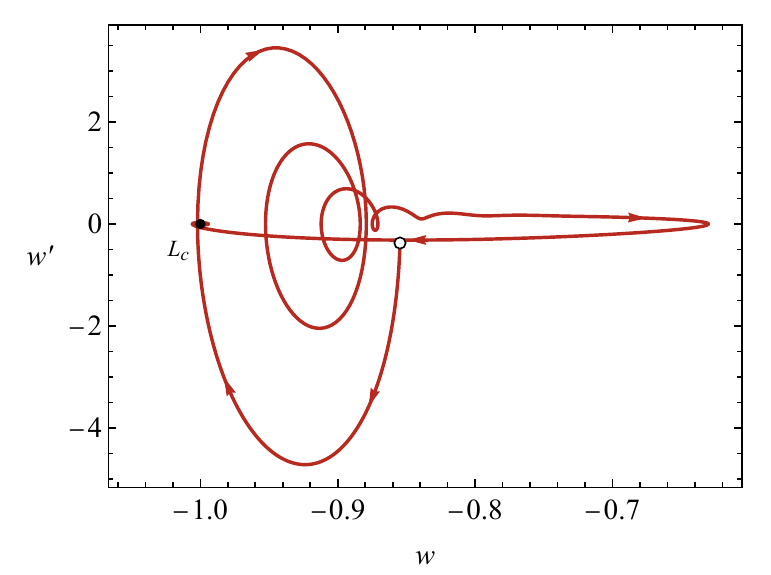}
     \caption{}
     \label{fig:w_wprime trajectory}
    \end{subfigure}
    \caption{A possible heteroclinic trajectory connecting the fixed points $M_{matter}$ and $L_c$, shown in both the $q-j$ plane (panel \ref{fig:q-j trajectory}) and the $w-w'$ plane (panel \ref{fig:w_wprime trajectory}). The trajectory in panel \ref{fig:q-j trajectory} is obtained as a solution of the system \eqref{DS_HS_1} and the one in panel \ref{fig:w_wprime trajectory} is obtained as a solution of the system \eqref{DS_HS_2}. The common initial conditions for both the panels are $\{r,m,m_2,m_3\}\vert_{\tilde{N}=0}=\{1.0013476935,0.000243363442,-2.0242206729,-2.9878932361\}$. For the left panel, the initial conditions on $q,\,j$ are taken as $\{q,j\}\vert_{\tilde{N}=0}=\{0.4979154952,1\}$, and the system in integrated forward up to $\tilde{N}=1518.6544$. For the right panel, the initial conditions on $\Omega_m,\,w,\,w'$ are taken as $\{\Omega_m,w,w'\}\vert_{\tilde{N}=0}=\{0.9983740863,-0.8547008547,-0.3725619110\}$, and the system is integrated forward until $\tilde{N}=447251.5961$. The initial conditions are chosen to be close to $M_{matter}$, but otherwise completely random and found purely by trial and error. Moreover, during the numerical exploration, it was also observed that the behaviour of a trajectory is highly sensitive to the initial conditions.}
    \label{fig:HS_trajectory}
\end{figure}

In a 6 or 7-dimensional phase space like \eqref{DS_HS_1} and \eqref{DS_HS_2}, there is no generic mathematical result guaranteeing the existence of heteroclinic trajectories connecting two (families of) fixed points. Consequently, one can only turn to numerical explorations to find possible trajectories. In Fig.\ref{fig:HS_trajectory}, we depict one such possible trajectory in both the $q-j$ plane and the $w-w'$ plane, to show both the kinematic and the dynamic evolution of the corresponding cosmology. The figure caption explicitly lists the initial conditions used to obtain this trajectory as solutions of the system \eqref{DS_HS_1} and \eqref{DS_HS_2}. These initial conditions are obtained purely by trial and error. It was also observed that the evolutions of trajectories are highly sensitive to the choice of the initial conditions, implying such trajectories are not generic.

\section{Application for a given $f(Q)$ model: $f(Q)=-2\Lambda+Q+\beta\sqrt{-Q}$}\label{sec:sqrt_f(Q)}

To showcase the applicability of the dynamical system formulation \eqref{DS_f(Q)_1}, let us turn towards the theory given by 
\begin{equation}\label{sqrt_model}
    f(Q)=-2\Lambda+Q+\beta\sqrt{-Q}\,.
\end{equation}
The version of this theory with $\Lambda=0$ has actually been very well-known since the popularization of the cosmological study of $f(Q)$ gravity. It is known to give rise to a cosmic evolution indistinguishable from $\Lambda$CDM \cite{BeltranJimenez:2019tme}, and it has been tested against available datasets in \cite{Atayde:2021pgb}. With $\Lambda=0$, one can calculate that
\begin{equation}
    m_0(Q)\equiv r(Q) = \frac{2\sqrt{-Q}-\beta}{2(\sqrt{-Q}-\beta)}\,, \qquad m_1(Q)\equiv m(Q) = \frac{\beta}{2(\sqrt{-Q}-\beta)}
\end{equation}
One can notice that the function $r=r(Q)$ is actually invertible to obtain $\sqrt{-Q}$ in terms of $r$, which, when substituted to the expression $m=m(Q)$, gives 
\begin{equation}
    m(r) = \frac{r - 1}{2r}\,.
\end{equation}
Substituting the above $m(r)$ into the $dr/dN$-equation of the system \eqref{DS_f(Q)_1} truncates the theory space hierarchy at the very first equation, and one is left with a 1-dimensional dynamical system. This is consistent with the argument provided in \cite{Boehmer:2021aji}, and later also confirmed in \cite{Dutta:2025fqw}, that the dynamical system of $f(Q)$ gravity in the presence of a single perfect fluid in the FLRW coincident gauge is 1-dimensional. 

The 1-dimensional cosmological phase space of the theory \eqref{sqrt_model} with a vanishing cosmological constant ($\Lambda=0$) can already be well investigated using the standard $f(Q)$ gravity dynamical system presented in \cite{Boehmer:2022wln} or \cite{Dutta:2025fqw}. Therefore in this paper, we show the merit of the dynamical system formulation \eqref{DS_f(Q)_1} by applying it for the case with a generic nonvanishing $\Lambda$. In fact it had earlier been pointed out that the generic $f(Q)$ model that gives rise to a cosmological evolution indistinguishable from the GR-$\Lambda$CDM model can have both a nonzero $\Lambda$ term and a nonzero $\sqrt{-Q}$ term \cite{Chakraborty:2025qlv}.

For a generic $\Lambda\neq0$, one has
\begin{equation}
    r(Q) = \frac{2Q + \beta\sqrt{-Q}}{2(-2\Lambda + Q + \beta\sqrt{-Q})}\,, \qquad m(Q) = \frac{\beta}{2(2\sqrt{-Q} - \beta)}\,.
\end{equation}
Note that the expression $r=r(Q)$ now leads to a quadratic equation for $\sqrt{-Q}$. If one still proceeds and solves the quadratic equation and then substitute it back into the expression for $m(Q)$, one ends up with a double-valued expression
\begin{equation}
    m(r) = \frac{\beta^2 \pm \beta\sqrt{\beta^2 - 4r(r-1)(8\Lambda - \beta^2)}}{4r(8\Lambda - \beta^2)}\,.
\end{equation}
Such a double-valued function $m(r)$ introduces a sort of ``branching'' in the dynamical system, and does not lead to a unique cosmological phase space. Consequently, the subsequent mathematical analysis will not be very clear. This was recently pointed out in \cite{Khyllep:2026pku}.

\begin{figure}[H]
    \centering
    \includegraphics[width=0.5\linewidth]{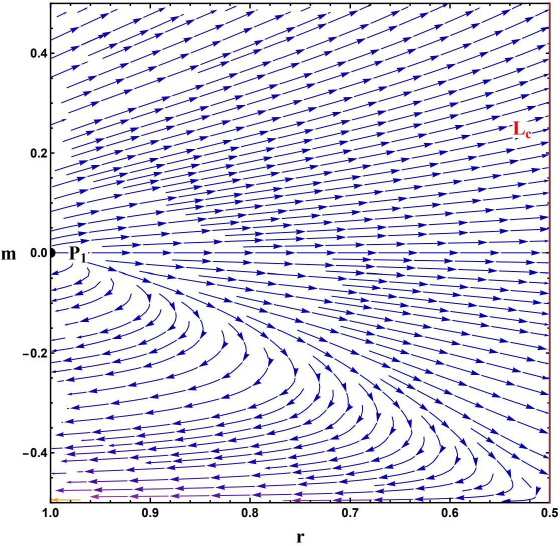}
    \caption{Cosmological phase space of the model $f(Q)=-2\Lambda+Q+\beta\sqrt{-Q}$ with $\Lambda,\beta\neq0$, as obtained from the system \eqref{DS_sqrt}}
    \label{fig:sqrt_f(Q)_portrait}
\end{figure}

However, one can note that for the theory \eqref{sqrt_model} with a generic $\Lambda\neq0$, $m_2=\frac{Qf_{QQQ}}{f_{QQ}}=-3/2$. Consequently, the hierarchy in Eq.\eqref{DS_f(Q)_1} collapses at the level of the $dm/dN$-equation, and we are left with the simple 2-dimensional system
\begin{subequations}\label{DS_sqrt}
    \begin{align}
    \frac{dr}{dN} &= 3(1 - r + m) \left( \frac{1 - 2r}{1 + 2m} \right) \, , \\
    \frac{dm}{dN} &= -\frac{3}{2} \frac{m}{r} (1-2r) \, .
\end{align}
\end{subequations}
It can be noted that now there is no more branching issues in the dynamical system, and it leads to a clear unique phase portrait. The physically viable fixed points (satisfying the condition \eqref{phys_viab_f(Q)_1}) of the system are listed in Table \ref{tab:sqrt_f(Q)_fp}.
\begin{table}[H]
    \centering
    \renewcommand{\arraystretch}{1.5}
    \begin{tabular}{l c l l}
        \hline
        \hline
        \textbf{Fixed Point} & \textbf{Coordinates \((r, m)\)} & \textbf{Stability} & \textbf{Cosmology} \\
        \hline
        \(P_1\) & \((1, 0)\) & Unstable node (Past attractor) & GR-matter-dominated epoch \\
        \(L_c\) & \((1/2, m)\) & Stable non-isolated attractor & Late-time de-Sitter epoch \\
        \hline
        \hline
    \end{tabular}
    \caption{Physically viable fixed points for the 2-dimensional dynamical system \eqref{DS_sqrt} corresponding to the model $f(Q)=-2\Lambda+Q+\beta\sqrt{-Q}$ with $\Lambda,\beta\neq0$.}
    \label{tab:sqrt_f(Q)_fp}
\end{table}
The phase portrait is shown in Fig.\ref{fig:sqrt_f(Q)_portrait}.

\section{Comparing $f(R)$ vs $f(Q)$ gravity theories that are \emph{kinematically} equivalent to $\Lambda$CDM}\label{sec:LCDM_mimicking}

Kinematical equivalence here means reproducing the same kind of cosmological evolution as the General Relativistic $\Lambda$CDM-model, which means obeying the cosmographic condition $j=1$ globally \cite{Dunajski:2008tg,Sahni:2002fz}. Kinematical equivalence does not necessarily mean dynamical equivalence ($w_{\rm DE}=-1$) \cite{Chakraborty:2022evc}. We term such theories as \emph{$\Lambda$CDM-mimicking} $f(R)$ or $f(Q)$ theories.

The condition of exactly reproducing a $\Lambda$CDM-like cosmic evolution is enforced by $j=1$, so that the next order cosmographic parameters snap ($s$) and lerk ($l$) can be expressed as
\begin{equation}
    s=-(2+3q)\,, \qquad l=6q^2+14q+9\,.
\end{equation}

\subsection{$\Lambda$CDM-mimicking $f(R)$ cosmology}\label{subsec:LCDM_fR}

Imposing the condition $j=1$ into the system \eqref{DS_f(R)_5} allows us to obtain a 3-dimensional system
\begin{subequations}\label{DS_LCDM_f(R)_1}
\begin{align}
\frac{\Omega_m}{f_R} &= 1 + (1-q)\left(\frac{1}{r}-1\right) - m\left(\frac{1+q}{1-q}\right)\geq0\,,
\\
\frac{dr}{dN} &= -r(1-r+m)\left(\frac{1+q}{1-q}\right)\,,
\\
\frac{dm}{dN} &= \frac{3(q-1)^2}{r(q+1)} + \frac{m(q^2+2q-1) - m^2(q+1)}{q-1} - \frac{(q-1)(q-2)}{q+1}
\\
\frac{dq}{dN} &= (2q-1)(q+1)\,.
\end{align}
\end{subequations}
The physically viable domain of the phase space is given by
\begin{equation}\label{phys_viab_LCDM_f(R)}
    {\cal D} = \left\lbrace(r,m,q)\in\mathbb{R}^3: -1\leq q\leq\frac{1}{2}\,\,\wedge\,\,m\geq0\,\,\wedge\,\,1 + (1-q)\left(\frac{1}{r}-1\right)\geq m\left(\frac{1+q}{1-q}\right)\right\rbrace\,,
\end{equation}
with the last inequality coming from the Friedmann constraint above.

The above system is singular at the de-Sitter limit $q\to-1$. In terms of the redefined time variable $\tilde{N}:d\tilde{N}=dN/(1+q)$, the dynamical system can be rewritten as
\begin{subequations}\label{DS_LCDM_f(R)_2}
\begin{align}
\frac{\Omega_m}{f_R} &= 1 + (1-q)\left(\frac{1}{r}-1\right) - m\left(\frac{1+q}{1-q}\right)\geq0\,,
\\
\frac{dr}{d\tilde{N}} &= -\frac{r(1-r+m)(1+q)^2}{1-q}\,,
\\
\frac{dm}{d\tilde{N}} &= \frac{3(q-1)^2}{r} - [m(q^2+2q-1) - m^2(q+1)]\left(\frac{1+q}{1-q}\right) - (q-1)(q-2)
\\
\frac{dq}{d\tilde{N}} &= (2q-1)(q+1)^2\,.
\end{align}
\end{subequations}

The physically viable fixed points calculated from the system \eqref{DS_LCDM_f(R)_2} are listed in Table \ref{tab:fR_LCDM_fixed_points}.
\begin{table}[H]
\centering
\renewcommand{\arraystretch}{1.7}
\begin{tabular}{l c c l l}
\hline\hline
Fixed Point & $(r, m, q)$ & Stability & Cosmology \\
\hline
$P_1$ & $\left(1, 0, \frac{1}{2}\right)$ & Saddle & GR-matter-dominated epoch 
\\
$P_2$ & $\left( \frac{7+\sqrt{73}}{12}, \frac{-5+\sqrt{73}}{12}, \frac{1}{2} \right)$ & Past attractor & Matter-\emph{like} cosmology \\
$L_c$ & $(2, m, -1)$ & Non-hyperbolic & Curvature dominated de-Sitter \\
\hline\hline
\end{tabular}
\caption{Physically viable fixed points for the dynamical system \eqref{DS_LCDM_f(R)_2} corresponding to $\Lambda$CDM-mimicking ($j=1$) $f(R)$ cosmology. $P_1$ represents the standard GR-matter-dominated epoch. $P_2$ represents a cosmological solution that is not necessarily General Relativistic or matter-dominated, but still has the matter-like evolution $a(t)\sim t^{2/3}$. $L_c$ is the line of curvature dominated de-Sitter attractors.}
\label{tab:fR_LCDM_fixed_points}
\end{table}
Interestingly, the past attractor of a $\Lambda$CDM-mimicking $f(R)$ cosmology is not the General Relativistic matter-dominated epoch ($P_1$), but rather a matter-\emph{like} cosmological solution $P_2$. For this cosmological solution, the gravity theory is not GR, and the contribution from the additional curvature fluid is significant ($m\neq0$); however, the curvature fluid as a whole acts like a perfect fluid with a vanishing equation of state parameter at the background level. This finding is in accordance with that in \cite{Chakraborty:2021jku}. The authors in \cite{Chakraborty:2021jku} studied $\Lambda$CDM-mimicking $f(R)$ cosmology using a different set of dynamical variables, and found that the GR-matter-dominated epoch is not a past attractor but actually a saddle.

\begin{figure}[H]
    \centering
    \includegraphics[width=\linewidth]{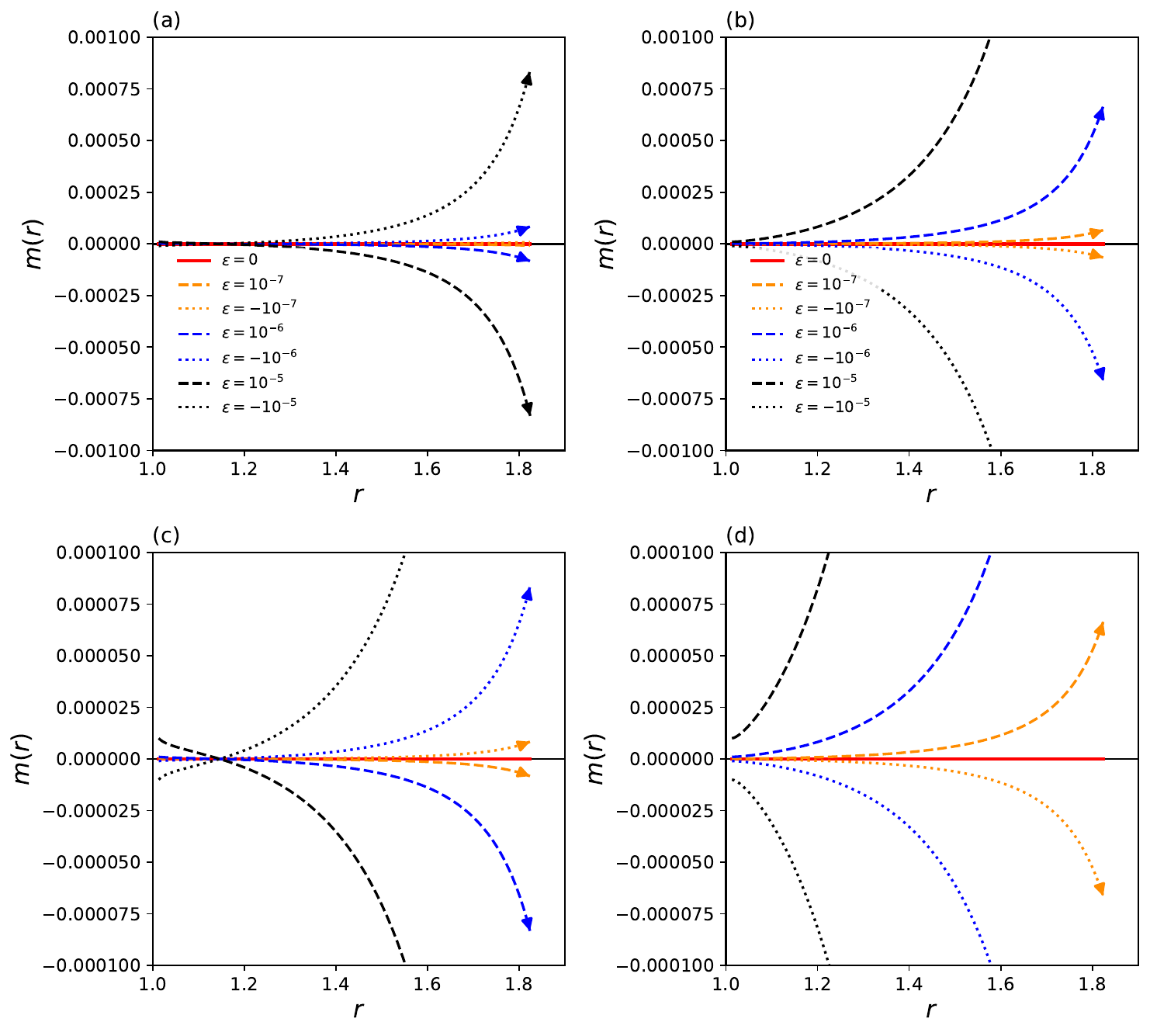}
    \caption{The dynamics of $\Lambda$CDM-mimicking $f(R)$ theories are shown as parametric curves ${r(q),m(q)}$ in the $m-r$ plane. The GR $\Lambda$CDM model ($m=0$, $r=(3q-3)/(q-2)$) is the central red line. Panels show numerical solutions of the nonautonomous system \eqref{eq:nonautonomous_LCDM_f(R)}. In Panel (a), trajectories start from the GR line shifted by $+\epsilon$ in both $r$ and $m$. In Panel (b), trajectories start from the GR line shifted by $-\epsilon$ in $r$ and $+\epsilon$ in $m$. Dashed/dotted orange, blue and black curves correspond to $\epsilon=\pm10^{-7},\pm10^{-6},\pm10^{-5}$ respectively. The evolution runs from $q=0.49$ to $q_0 \approx -0.55$ (today). Panel (c) and (d) are zoomed-in version of the figure (a) and (b), respectively. Zoomed panels reveal that GR is not a generic past attractor, in accordance with the results obtained in Table \ref{tab:fR_LCDM_fixed_points}. Also, a solution can start in the physically viable region ($m>0$) and cross into the region plagued by theoretical pathology ($m<0$).}
    \label{fig:LCDM_mimicking_m(r)_f(R)}
\end{figure}

It is perhaps more illuminating to visualize the $\Lambda$CDM-mimicking $f(R)$ cosmological solutions as flows in the theory space $r-m$. To do that, it is useful to write the dynamical system \eqref{DS_LCDM_f(R)_1} in the non-autonomous form
\begin{subequations}\label{eq:nonautonomous_LCDM_f(R)}
    \begin{eqnarray}
        && \frac{dr}{d(-q)} = - \frac{dr/dN}{dq/dN} = \frac{r(1-r+m)}{(1-q)(2q-1)}\,,\nonumber
        \\
        && \frac{dm}{d(-q)} = - \frac{dm/dN}{dq/dN} = -\frac{r \left[m^2 (q+1)^2 - m(q+1) \left(q^2+2q-1\right) - (2-q)(1-q)^2\right]+3 (1-q)^3}{(q+1)^2 (1-q) (2q-1) r}\,.\nonumber
        \\
        &&\nonumber
    \end{eqnarray}
\end{subequations}
The above system can also be obtained directly by substituting $j=1,\,s=-2-3q$ into the $j(q)$-closed non-autonomous system \eqref{DS_f(R)_j(q)}. The non-autonomous system \eqref{eq:nonautonomous_LCDM_f(R)} has recently been utilized in \cite{Chakraborty:2025lkz} to investigate the behaviour of $\Lambda$CDM-mimicking $f(R)$ theories in the theory space $r-m$. More precisely, when the solutions of the non-autonomous system \eqref{eq:nonautonomous_LCDM_f(R)} are portrayed in the $r-m$ plane, they show how the deviation from the underlying theory from GR evolves along the course of the cosmic evolution. Following the Ref.~\cite{Chakraborty:2025lkz}, we show some such solution curves in Fig.\ref{fig:LCDM_mimicking_m(r)_f(R)}, where we set the initial conditions such that at $q=0.49$ the underlying $f(R)$ theory is very close to GR, with small deviations from GR parametrized by a smallness parameter $\epsilon$. The exact initial conditions taken to produce these curves are specified in the caption.

In Fig.\ref{fig:LCDM_mimicking_m(r)_f(R)}, the parametric plots in the panels (a) and (b) give an apparent impression that GR acts as the cosmological past attractor for the $\Lambda$CDM-mimicking $f(R)$ theories; that all such $f(R)$ theories asymptote to GR in the asymptotic past. However, we emphasize that there is no such generic tendency. This is made explicit in the panels (c) and (d), which are zoomed-in versions of the panels (a) and (b), respectively. Although the parametric plots for solutions corresponding to the initial condition $\lbrace r(0.49),m(0.49) \rbrace = \lbrace \frac{3q-3}{q-2}\vert_{q=0.49} - \epsilon,\epsilon\rbrace$ do seem to asymptote to GR in the far past (panel (d)), this behaviour is absent for solutions corresponding to the initial condition $\lbrace r(0.49),m(0.49) \rbrace = \lbrace \frac{3q-3}{q-2}\vert_{q=0.49} + \epsilon,\epsilon\rbrace$ (panel (c)). Rather, for the solutions depicted in panel (c), it appears that the underlying theory actually starts deviating from GR also in the past. This finding is in accordance with the mathematical result obtained in Table \ref{tab:fR_LCDM_fixed_points} that the GR-matter dominated epoch $P_1$ is actually a saddle and not a past attractor.

The solution trajectories with the initial conditions $\lbrace r(0.49),m(0.49) \rbrace = \lbrace \frac{3q-3}{q-2}\vert_{q=0.49} + \epsilon,\epsilon\rbrace$ also show that a $\Lambda$CDM-mimicking $f(R)$ cosmology can start as a physically healthy theory near the matter-dominated epoch but end up as one plagued by either ghost or tachyonic instability (i.e., the solution curve switching from the $m>0$ to $m<0$ region).

\subsection{$\Lambda$CDM-mimicking $f(Q)$ cosmology}\label{subsec:LCDM_fQ}

Substituting $j=1$ directly into \eqref{eq:fQ_dmdN_qjm} gives a remarkably simple closed 2-dimensional system
\begin{subequations}\label{DS_LCDM_f(Q)}
\begin{align}
\frac{\Omega_m}{f_Q} &= \frac{2}{3}(1+2m)(1+q)\geq0\,,
\\
\frac{dm}{dN} &= m(1+2m)(1+q)\,,
\\
\frac{dq}{dN} &= (2q-1)(q+1)\,.
\end{align}
\end{subequations}
The physically viable domain of the phase space is
\begin{equation}\label{phys_viab_LCDM_f(Q)}
    {\cal D} = \{(m,q)\in\mathbb{R}^2: -1\leq q\leq1/2,\,m\geq-1/2\}\,.
\end{equation}

Unlike its $f(R)$ counterpart \eqref{DS_LCDM_f(R)_1}, this is actually not singular at the de-Sitter limit $q\to-1$. Hence, no redefinition of the time variable is necessary, and we can perform a fixed point analysis directly on the system \eqref{DS_LCDM_f(Q)}. The physically viable fixed points calculated from the system \eqref{DS_LCDM_f(Q)} are listed in Table \ref{tab:fQ_LCDM_fixed_points}.
\begin{table}[H]
\centering
\renewcommand{\arraystretch}{1.7}
\begin{tabular}{l c c l l}
\hline\hline
Fixed Point & $(m, q)$ & Stability & Cosmology \\
\hline
$P_1$ & $\left(0, \frac{1}{2}\right)$ & past attractor & GR-matter-dominated epoch
\\
$P_2$ & $\left(-\frac{1}{2}, \frac{1}{2}\right)$ & Saddle & Matter-\emph{like} cosmology 
\\
$L_c$ & $(m, -1)$ & Future attractor & Nonmetricity dominated de-Sitter \\
\hline\hline
\end{tabular}
\caption{Physically viable fixed points for the dynamical system \eqref{DS_LCDM_f(Q)} corresponding to $\Lambda$CDM-mimicking ($j=1$) $f(Q)$ cosmology. $P_1$ represents the standard GR-matter-dominated epoch. $P_2$ represents a cosmological solution that is not necessarily General Relativistic or matter-dominated, but still has the matter-like evolution $a(t)\sim t^{2/3}$. $L_c$ is the line of nonmetricity dominated de Sitter attractors.}
\label{tab:fQ_LCDM_fixed_points}
\end{table}

\begin{figure}[H]
    \centering
    \includegraphics[width=0.5\linewidth]{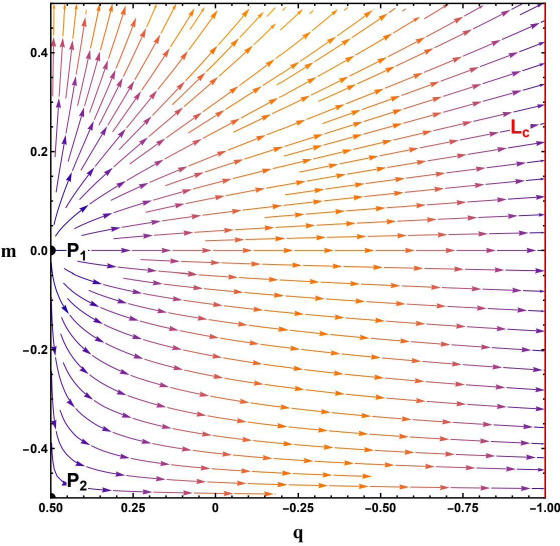}
    \caption{Cosmological phase space of $\Lambda$CDM-mimicking $f(Q)$ gravity, as obtained from the system \eqref{DS_LCDM_f(Q)}.}
    \label{fig:LCDM-mimicking_m(q)_f(Q)}
\end{figure}

It is worth pointing out the crucial differences between the fixed point structures of $\Lambda$CDM-mimicking $f(Q)$, as we obtain in Table \ref{tab:fQ_LCDM_fixed_points}, and $\Lambda$CDM-mimicking $f(R)$, as we had earlier obtained in table \ref{tab:fR_LCDM_fixed_points}. Apart from the mathematical difference that the cosmological phase space of $\Lambda$CDM-mimicking $f(Q)$ is 2-dimensional, as opposed to the 3-dimensional phase space of $\Lambda$CDM-mimicking $f(R)$, the physical difference is that the General Relativistic matter dominated epoch $P_1$ does in fact appear as a past attractor. This can also be clearly seen from the 2-dimensional phase portrait presented in Fig.\ref{fig:LCDM-mimicking_m(q)_f(Q)}. In other words, for $\Lambda$CDM-mimicking $f(Q)$ cosmologies, GR act as a genuine cosmological past attractor.

It is worth pointing out that, from the simple closed 2-dimensional system \eqref{DS_LCDM_f(Q)}, one can write
\begin{equation}
    \frac{dm}{dq} = -\frac{m(1+2m)}{1-2q}\,,
\end{equation}
which can actually be analytically solved as
\begin{equation}\label{eq:f(Q)_LCDM_exact}
    m(q) = \frac{C\sqrt{1-2q}}{1-2C\sqrt{1-2q}}\,, \qquad\qquad C=\frac{3\Omega_{m0}-2f_{Q0}(1+q_0)}{6\Omega_{m0}\sqrt{1-2q_0}}\,,
\end{equation}
where the constant $C$ is calculated by requiring that at $q=q_0$, one has $\{\Omega_m,f_Q\}=\{\Omega_{m0},f_{Q0}\}$, and utilizing the Friedmann equation \eqref{fried-f(Q)} and the Raychaudhuri equation \eqref{raych-f(Q)}. Very recently, it has been discovered that the mathematical system within the $f(Q)$ framework that exactly reproduces a $\Lambda$CDM-like cosmic evolution $j=1$ is actually exactly solvable \cite{Khyllep:2026pku}. Our finding here re-establishes and reinforces that result using the variables $m$ and $q$.

\section{Comparing $f(R)$ vs $f(Q)$ gravity theories that are \emph{dynamically} equivalent to $\Lambda$CDM}\label{sec:cosm_const_mimicking}

The kinematical equivalence condition $j=1$ studied in Sec.~\ref{sec:LCDM_mimicking} constrains only the expansion history $a(t)$, but it does not by itself force $w=-1$. There is a difference between kinematical and dynamical equivalence with $\Lambda$CDM, characterized by the condition $j=1$ and $w=-1$, respectively \cite{Chakraborty:2022evc}. It is perfectly possible to reproduce the $\Lambda$CDM-like background evolution while the underlying dark energy fluid has a genuinely evolving equation of state. In this section we instead impose the strictly stronger condition of \emph{dynamical} equivalence to $\Lambda$CDM, namely $w=-1$ identically throughout the cosmic evolution, directly on the equation-of-state-closed systems \eqref{DS_f(R)_6} and \eqref{DS_f(Q)_3} of Secs.~\ref{subsec:eos_fR} and \ref{subsec:eos_fQ}. Physically, the additional dynamical degree of freedom constituted by the curvature or the nonmetricity fluid of Eqs.\eqref{curv_fluid} or \eqref{nonmet_fluid} must evolve strictly in a way that the corresponding equation of state parameters \eqref{DE_eos_f(R)} and \eqref{DE_eos_f(Q)} remains constant at a value $-1$. We refer to such theories as \emph{cosmological constant mimicking} $f(R)$ or $f(Q)$ theories. 

\subsection{Cosmological constant mimicking $f(R)$ cosmology}\label{subsec:cosm_const_fR}

Setting $w=w_c=-1$ (hence $w'=w''=0$) in the quantities $A,\,B,\,P$ of \eqref{DS_f(R)_6} gives
\begin{equation}
    A = 4-3\Omega_m\,,\qquad B=\Omega_m\,,\qquad P = \Omega_m\left(9\Omega_m^2-8\right)\,.
\end{equation}
The $w$ and $w'$ equations of \eqref{DS_f(R)_6} become trivial, and the system closes on the three variables $\{\Omega_m,r,m\}$:
\begin{subequations}\label{DS_wm1_f(R)_1}
\begin{align}
    \frac{\Omega_m}{f_R} &= \frac{(4-3\Omega_m)^2-r\left[6\Omega_m m+9(\Omega_m-1)^2-1\right]}{2r(4-3\Omega_m)}\,,
    \\
    \frac{d\Omega_m}{dN} &= -3\Omega_m(1-\Omega_m)\,,
    \\
    \frac{dr}{dN} &= -\frac{3\Omega_m\,r(1-r+m)}{4-3\Omega_m}\,,
    \\
    \frac{dm}{dN} &= \frac{-rmP+6rm^2B^2+A^2\left[(r-3)(\Omega_m-1)-r+1\right]}{2rAB}\,.
\end{align}
\end{subequations}
The physically viable domain of the phase space is given by
\begin{equation}\label{phys_viab_wm1_f(R)}
    {\cal D} = \left\lbrace(\Omega_m,r,m)\in\mathbb{R}^3: 0\leq\Omega_m\leq1\,\,\wedge\,\,m\geq0\,\,\wedge\,\,\frac{(4-3\Omega_m)^2-r\left[6\Omega_mm+9(\Omega_m-1)^2-1\right]}{2r(4-3\Omega_m)}\geq0\right\rbrace\,,
\end{equation}
with the last inequality coming from the Friedmann constraint above.

As with its $j=1$ counterpart \eqref{DS_LCDM_f(R)_1}, the system \eqref{DS_wm1_f(R)_1} is singular at the de Sitter limit $\Omega_m\to0$, where $B=\Omega_m$ vanishes in the denominator of $dm/dN$. Confining ourselves to the physically viable region, we regularize with the time redefinition
\begin{equation}
    dN\to d\tilde{N}=\frac{dN}{\Omega_m}\,,
\end{equation}
in terms of which \eqref{DS_wm1_f(R)_1} becomes
\begin{subequations}\label{DS_wm1_f(R)_2}
\begin{align}
    \frac{d\Omega_m}{d\tilde{N}} &= -3\Omega_m^2(1-\Omega_m)\,,
    \\
    \frac{dr}{d\tilde{N}} &= -\frac{3\Omega_m^2\,r(1-r+m)}{4-3\Omega_m}\,,
    \\
    \frac{dm}{d\tilde{N}} &= \frac{-rmP+6rm^2B^2+A^2\left[(r-3)(\Omega_m-1)-r+1\right]}{2rA}\,,
\end{align}
\end{subequations}
which is now regular throughout $0\le\Omega_m\le1$.

The physically viable fixed points of \eqref{DS_wm1_f(R)_2} are listed in Table~\ref{tab:fR_wm1_fixed_points}.
\begin{table}[H]
\centering
\renewcommand{\arraystretch}{1.7}
\begin{tabular}{l c c l l}
\hline\hline
Fixed Point & $(\Omega_m, r, m)$ & Stability & Cosmology \\
\hline
$P_1$ & $\left(1,\,1,\,0\right)$ & Saddle & GR-matter-dominated epoch
\\
$P_2$ & $\left(1,\, \frac{7+\sqrt{73}}{12},\, \frac{-5+\sqrt{73}}{12} \right)$ & Past attractor & \emph{Gravity-decoupled} matter-dominated solution \\
$L_c$ & $(0,\,2,\,m)$ & Non-hyperbolic & Curvature dominated de-Sitter \\
\hline\hline
\end{tabular}
\caption{Physically viable fixed points for the dynamical system \eqref{DS_wm1_f(R)_2} corresponding to $w=-1$-mimicking $f(R)$ cosmology. $P_1$ represents the standard GR-matter-dominated epoch. $P_2$ represents a cosmological solution that is matter-dominated, but with vanishing effective subhorizon gravitational coupling ($1/f_R~\to0$) between matter and geometry. $L_c$ is the line of curvature dominated de-Sitter attractors.}
\label{tab:fR_wm1_fixed_points}
\end{table}

\begin{figure}[H]
    \centering
    \includegraphics[width=\linewidth]{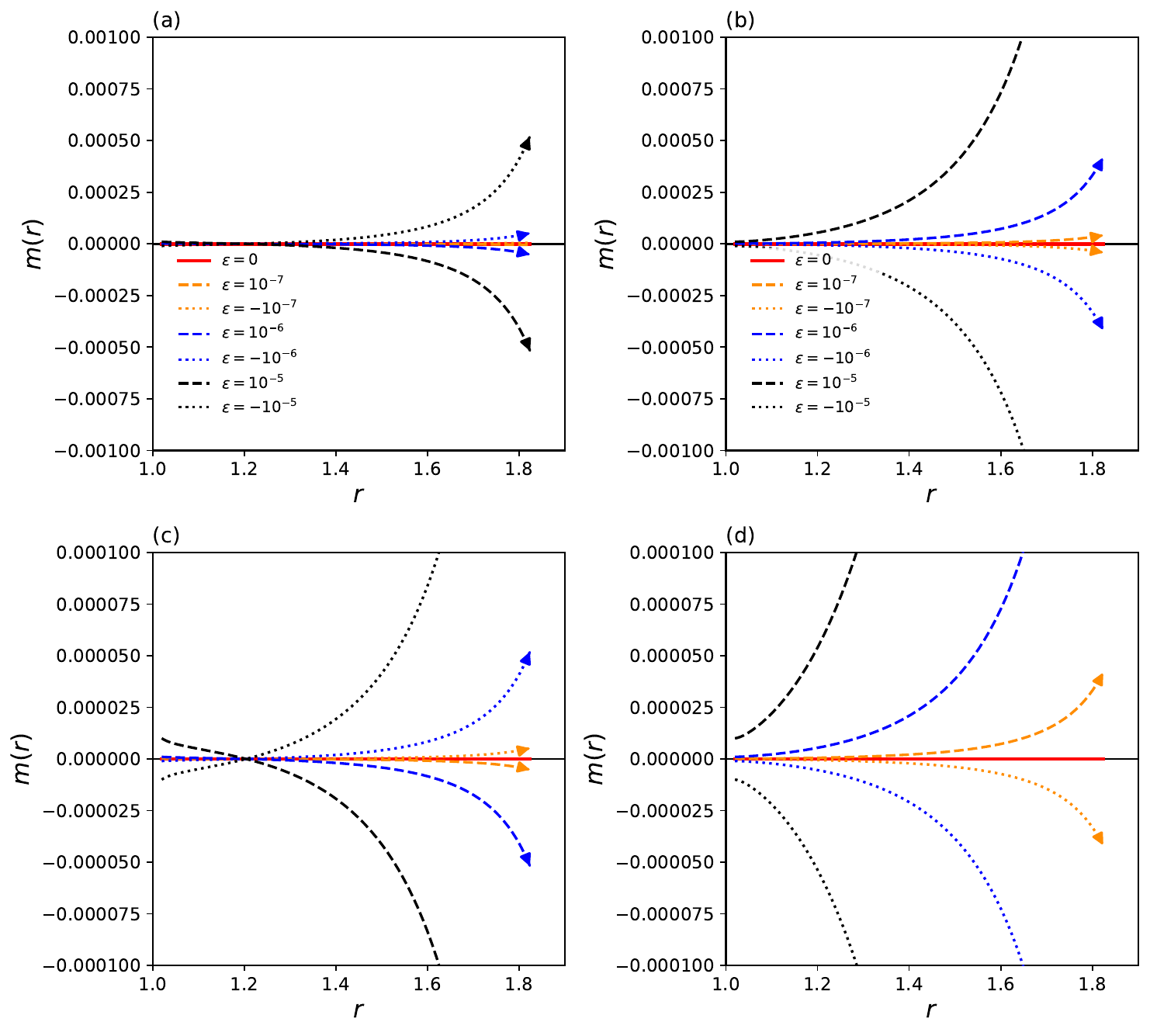}
    \caption{The dynamics of $w=-1$-cosmological constant mimicking $f(R)$ theories, shown as parametric curves $\{r(\Omega_m),m(\Omega_m)\}$ in the $m$-$r$ plane, obtained as numerical solutions of the non-autonomous system \eqref{eq:nonautonomous_wm1_f(R)}. The GR $\Lambda$CDM model ($m=0$, $r=r_{\rm GR}(\Omega_m)=(3\Omega_m-4)/(\Omega_m-2)$) is the central red line. In panel (a), trajectories start from the GR line at $\Omega_m=0.99$ shifted by $+\epsilon$ in both $r$ and $m$. In panel (b), trajectories start from the GR line shifted by $-\epsilon$ in $r$ and $+\epsilon$ in $m$. Dashed/dotted orange, blue and black curves correspond to $\epsilon=\pm10^{-7},\pm10^{-6},\pm10^{-5}$ respectively. The evolution runs from $\Omega=0.99$ to $\Omega_m \approx 0.3$ (today).  Panels (c) and (d) are zoomed-in versions of (a) and (b). As for the $j=1$ case, GR is not a generic past attractor, in accordance with the results obtained in Table \ref{tab:fR_wm1_fixed_points}. A solution can start in the physically viable region ($m>0$) and cross into the region plagued by theoretical pathologies ($m<0$).}
    \label{fig:wm1_mr_fR}
\end{figure}

As in Sec.~\ref{subsec:LCDM_fR}, it is more illuminating to visualize the cosmological constant mimicking $f(R)$ solutions as flows in the theory space $r$-$m$, using $-\Omega_m$ (which, like $-q$ before, is monotonically increasing at late times) as a proxy time variable. In order to do that, we rewrite the dynamical system \eqref{DS_wm1_f(R)_2} in the following non-autonomous form
\begin{subequations}\label{eq:nonautonomous_wm1_f(R)}
\begin{align}
    \frac{dr}{d(-\Omega_m)} &= -\frac{r(1-r+m)}{(1-\Omega_m)(4-3\Omega_m)}\,,
    \\
    \frac{dm}{d(-\Omega_m)} &= \frac{-rmP+6rm^2B^2+A^2\left[(r-3)(\Omega_m-1)-r+1\right]}{6rAB\,\Omega_m(1-\Omega_m)}\,,
\end{align}
\end{subequations}
with $A,\,B,\,P$ as above. The above system can also be obtained directly by substituting $w_c=-1$ into the constant-equation-of-state system \eqref{DS_f(R)_wc}. It dictates the flow of the solution curves in the theory space $r$-$m$ for $f(R)$ theories whose curvature fluid behaves as a whole exactly like a cosmological constant. We show some such solution curves in Fig.~\ref{fig:wm1_mr_fR}, with initial conditions set at $\Omega_m=0.99$, evolved forward to $\Omega_m=0.3$.

\subsection{Cosmological constant mimicking $f(Q)$ cosmology}\label{subsec:cosm_const_fQ}

Setting $w=-1$  closes on the system \eqref{DS_f(Q)_3} on two variables $\{\Omega_m,m\}$:
\begin{subequations}\label{DS_wm1_f(Q)}
\begin{align}
    f_Q &= \frac{1}{1+2m}\,,
    \\
    \frac{d\Omega_m}{dN} &= -3\Omega_m(1-\Omega_m)\,,
    \\
    \frac{dm}{dN} &= \frac{3}{2}m\Omega_m(1+2m)\,. 
\end{align}
\end{subequations}
The physically viable domain of the phase space is
\begin{equation}\label{phys_viab_wm1_f(Q)}
    {\cal D} = \{(\Omega_m,m)\in\mathbb{R}^2: 0\leq\Omega_m\leq1,\,m>-1/2\}\,.
\end{equation}

\begin{figure}[H]
    \centering
    \includegraphics[width=0.5\linewidth]{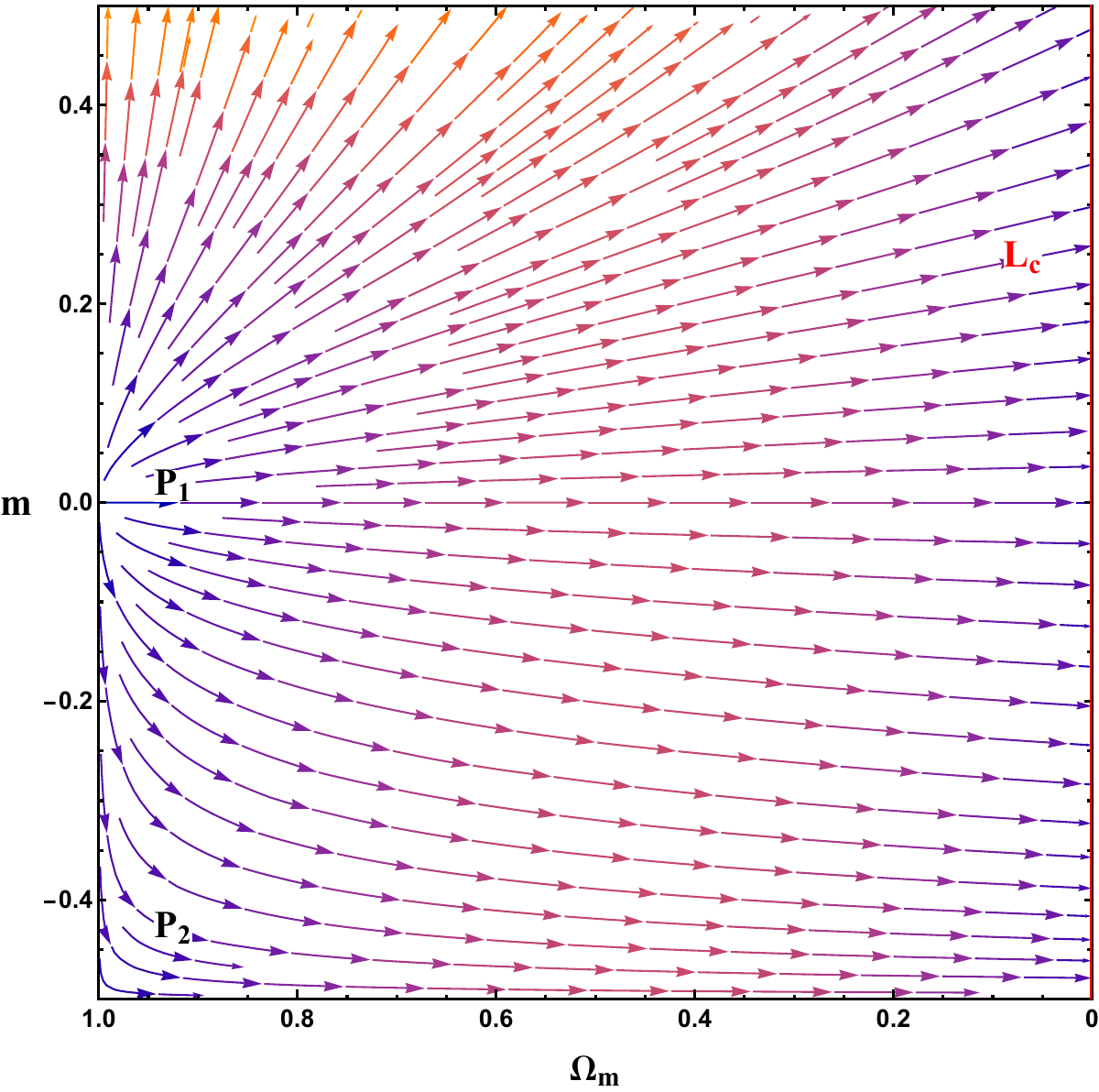}
    \caption{Cosmological phase space of cosmological constant mimicking $f(Q)$ gravity, as obtained from the system \eqref{DS_wm1_f(Q)}. The phase portrait perfectly mirrors Fig.~\ref{fig:LCDM-mimicking_m(q)_f(Q)}.}
    \label{fig:wm1_mOm_fQ}
\end{figure}

Unlike the $f(R)$ case, the system \eqref{DS_wm1_f(Q)} is not singular at $\Omega_m\to0$, and we can perform a fixed-point analysis directly on \eqref{DS_wm1_f(Q)} without any time redefinition, exactly as for the $j=1$ case. The physically viable fixed points are listed in Table~\ref{tab:fQ_wm1_fixed_points}.
\begin{table}[H]
\centering
\renewcommand{\arraystretch}{1.7}
\begin{tabular}{l c c l l}
\hline\hline
Fixed Point & $(\Omega_m, m)$ & Stability & Cosmology \\
\hline
$P_1$ & $\left(1,\,0\right)$ & Past attractor & GR-matter-dominated epoch
\\
$P_2$ & $\left(1,\,-\frac{1}{2}\right)$ & Saddle & \emph{Gravity-decoupled} matter dominated solution
\\
$L_c$ & $(0,\,m)$ & Future attractor & Nonmetricity dominated de-Sitter \\
\hline\hline
\end{tabular}
\caption{Physically viable fixed points for the dynamical system \eqref{DS_wm1_f(Q)} corresponding to cosmological constant mimicking ($w=-1$) $f(Q)$ cosmology. $P_1$ represents the standard GR-matter-dominated epoch. $P_2$ represents a cosmological solution that is matter-dominated, but with vanishing effective subhorizon gravitational coupling ($1/f_Q~\to0$) between matter and geometry. $L_c$ is the line of nonmetricity dominated de-Sitter attractors.}
\label{tab:fQ_wm1_fixed_points}
\end{table}
As we see again, unlike $f(R)$, the General Relativistic matter-dominated epoch $P_1$ is a genuine cosmological past attractor: for cosmological constant mimicking $f(Q)$ cosmologies, GR is approached generically in the past. The full phase portrait is shown in Fig.~\ref{fig:wm1_mOm_fQ}.

Finally, exactly as for the $j=1$ case, the closed system \eqref{DS_wm1_f(Q)} is analytically solvable. Eliminating $N$ gives
\begin{equation}
    \frac{dm}{d\Omega_m} = -\frac{m(1+2m)}{2(1-\Omega_m)}\,,
\end{equation}
which separates and integrates to
\begin{equation}\label{eq:f(Q)_wm1_exact}
    m(\Omega_m) = \frac{C\sqrt{1-\Omega_m}}{1-2C\sqrt{1-\Omega_m}}\,,\qquad\qquad C = \frac{1-f_{Q0}}{2\sqrt{1-\Omega_{m0}}}\,,
\end{equation}
where the integration constant $C$ is fixed by requiring $\{\Omega_m,f_Q\}=\{\Omega_{m0},f_{Q0}\}$ at $\Omega_m=\Omega_{m0}$, using the Friedmann equation above. Equation~\eqref{eq:f(Q)_wm1_exact} is, as it must be, precisely the $w=-1$ specialization of the exact $j=1$ solution found in Eq.\eqref{eq:f(Q)_LCDM_exact} (also obtained in \cite{Khyllep:2026pku}); the former can be found from the latter by substituting $w=-1\Leftrightarrow q_0=-1+3\Omega_{m0}/2$.

\section{A discussion on the phase space of kinematically vs dynamically $\Lambda$CDM-mimicking $f(R)$ and $f(Q)$ cosmology}\label{sec:kin_vs_dyn}

Setting $w=-1$ identically in the statefinder relation \eqref{j} (in other words, setting $w=-1,w'=0$) gives $j=1$ identically. Dynamical equivalence to $\Lambda$CDM automatically implies kinematic equivalence. However, the converse is not true. The solution $w(z)=-1$ is a measure-zero solution of all the possibilities admitted by $j=1$ \cite{Chakraborty:2025rvc}. The general solution to $j=1$ is given by the equation $w'=3w(1+w)$, as seen by substituting $j=1$ in the relation \eqref{statefinder_s}. In that sense, the phase space of the \emph{cosmological constant} mimicking ($w=-1$) $f(X)$ theories must be a subspace of the entire phase space of $\Lambda$CDM-mimicking ($j=1$) $f(X)$ theories. This is a very subtle and crucial point that deserves elaboration -- and we attend to it in this section. 

At first sight, a side-by-side comparison of Secs.~\ref{sec:LCDM_mimicking} and \ref{sec:cosm_const_mimicking} seems to sit uneasily with this expectation. On one hand, the reduced dynamical systems of the two sections produce identical theory-space flows (compare Figs.~\ref{fig:LCDM_mimicking_m(r)_f(R)} and \ref{fig:wm1_mr_fR}) and identical phase portraits (compare Figs.~\ref{fig:LCDM-mimicking_m(q)_f(Q)} and \ref{fig:wm1_mOm_fQ}), with fixed points sharing the same theory-space coordinates and the same stability character. On the other hand, the fixed point $P_2$ admits two different physical readings for the two different closures. In Tables~\ref{tab:fR_LCDM_fixed_points} and \ref{tab:fQ_LCDM_fixed_points} it is a matter-\emph{like} solution for which $\Omega_m/f_X=0$ is realized by $\Omega_m=0$, and for which the statefinder relation \eqref{q} with $q=1/2$ then forces $w=0$; this is precisely the asymptotic past value of the generic solution of $w'=3w(1+w)$ for $j=1$. In Tables~\ref{tab:fR_wm1_fixed_points} and \ref{tab:fQ_wm1_fixed_points}, instead, $P_2$ sits at $\Omega_m=1$, so that $\Omega_m/f_X=0$ can only be realized by $1/f_X\to0$, which can be interpreted as a limit at which gravity decouples from matter (considering that $1/f_x$ is the effective gravitational coupling at the subhorizon limit). If the $w=-1$ phase space is a subspace of the $j=1$ phase space, where in the list of $j=1$ fixed points is the gravity-decoupled $P_2$ of Sec.~\ref{sec:cosm_const_mimicking}? 

The resolution, which we now present, lies in the deeper realization that the phase space of the cosmographic closure is, in reality, one dimension larger than what the reduced systems \eqref{DS_LCDM_f(R)_1} and \eqref{DS_LCDM_f(Q)} suggest.

\subsection{The \emph{lifted} phase space for cosmographic closure: the \emph{base} and the \emph{fibre} dimension}\label{subsec:fibre}

One can note that the Friedmann constraint determines only the ratio $\Omega_m/f_X$, and neither $\Omega_m$ nor $f_X$ separately. However, independently of the underlying theory and of any closure, the matter density parameter obeys a purely kinematic evolution equation
\begin{equation}\label{eq:Om_kinematic}
    \frac{d\Omega_m}{dN}=(2q-1)\,\Omega_m\,,
\end{equation}
$\Omega_m$ itself never enters the evolution equations of the cosmographic-closure systems \eqref{DS_LCDM_f(R)_1} and \eqref{DS_LCDM_f(Q)}; those systems close on $\{r,m,q\}$ and $\{m,q\}$, without ever needing $\Omega_m$. Furthermore, its evolution is entirely determined by the cosmographic parameter $q$. The complete phase space of $\Lambda$CDM-mimicking cosmology is, in reality, the \emph{lifted} space obtained by appending Eq.~\eqref{eq:Om_kinematic} to \eqref{DS_LCDM_f(R)_1} or \eqref{DS_LCDM_f(Q)}. It is 4-dimensional, spanned by $\{r,m,q,\Omega_m\}$, for $f(R)$ gravity, and 3-dimensional, spanned by $\{m,q,\Omega_m\}$, for $f(Q)$ gravity. $f_X$ is then read off from the Friedmann constraint once $\Omega_m$ is known. The lifted system has a skew-product structure: the \emph{base} $\{r,m,q\}$ or $\{m,q\}$ evolves autonomously, while the \emph{fibre} coordinate $\Omega_m$ is driven by the base. The reduced systems of Sec.~\ref{sec:LCDM_mimicking} are the projection of the lifted system onto the base.

\subsection{The cosmological constant mimicking phase space as an invariant leaf}\label{subsec:leaf}

\begin{figure}[!h]
    \centering
    \includegraphics[width=0.78\linewidth]{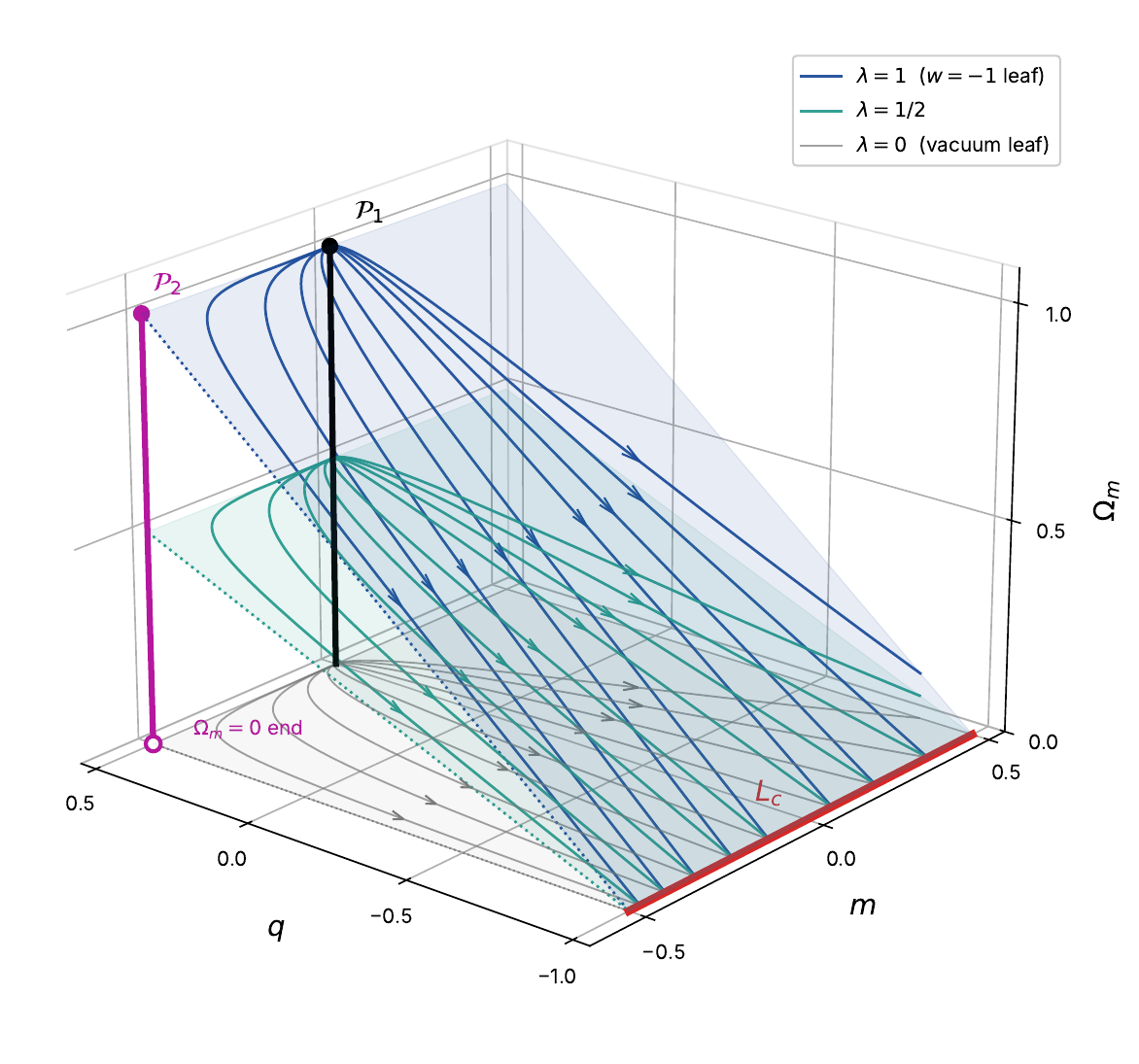}
    \caption{The lifted phase space $\{m,q,\Omega_m\}$ of $\Lambda$CDM-mimicking $f(Q)$ cosmology. The shaded planes are the invariant leaves $\Sigma_\lambda$ of Eq.~\eqref{eq:leaves} for $\lambda=1$ (the cosmological constant mimicking phase space of Sec.~\ref{subsec:cosm_const_fQ}), $\lambda=1/2$ and $\lambda=0$ (the vacuum leaf). All leaves meet along the de Sitter line $L_c$ (red). The trajectories drawn on each leaf are the exact solutions \eqref{eq:f(Q)_LCDM_exact}, and the dotted line on each leaf is the invariant line $m=-1/2$. Every leaf projects onto the base $\{m,q\}$ as an identical copy of Fig.~\ref{fig:LCDM-mimicking_m(q)_f(Q)}. The vertical lines are the lines of fixed points $\mathcal{P}_1$ (black) and $\mathcal{P}_2$ (magenta) of Table~\ref{tab:lifted_fp}, which cross each leaf $\Sigma_\lambda$ at $\Omega_m=\lambda$. The $w=-1$ leaf meets $\mathcal{P}_2$ at its upper end $\Omega_m=1$ (filled circle), whereas the matter-like $P_2$ of Sec.~\ref{subsec:LCDM_fQ} corresponds to its lower end $\Omega_m=0$ (open circle).}
    \label{fig:lifted_fQ}
\end{figure}

For the $j=1$ closure, $dq/dN=(2q-1)(q+1)$, and Eq.~\eqref{eq:Om_kinematic} immediately implies that the quantity $\lambda\equiv\frac{3\Omega_m}{2(1+q)}$ is a first integral of the lifted system:
\begin{equation}\label{eq:lambda}
    \lambda\equiv\frac{3\Omega_m}{2(1+q)}\,\Rightarrow\,\frac{d\lambda}{dN}=0\,.
\end{equation}
The larger lifted phase space can therefore imagined to be foliated by a family of invariant hypersurfaces (leaves)
\begin{equation}\label{eq:leaves}
    \Sigma_\lambda\equiv\left\lbrace\Omega_m-\frac{2}{3}\lambda(1+q)=0\right\rbrace\,,\qquad 0\leq\lambda\leq1\,,
\end{equation}
the upper bound following from $\Omega_m\leq1$ at $q=1/2$. All the leaves meet along the de Sitter boundary $\{q=-1,\,\Omega_m=0\}$, which contains the line $L_c$, and fan out from it. For $f(Q)$ gravity, where the lifted space is three-dimensional, this structure can be visualized directly; see Fig.~\ref{fig:lifted_fQ}.

The first integral $\lambda$ has a transparent physical meaning: it is the fraction of the $a^{-3}$-component of the expansion rate that is supplied by genuine nonrelativistic matter. To see this, note that $j=1$ gives rise to a $\Lambda$CDM-like expansion, one has
\begin{equation}
   3H^2=Aa^{-3}+B, \qquad  \frac{2}{3}(1+q)H^2=Aa^{-3}, \qquad A,B\geq0
\end{equation}
$3H^2=Aa^{-3}+B$, with constants $A,B\geq0$, for which $\frac{2}{3}(1+q)=Aa^{-3}/(Aa^{-3}+B)$. On the leaf $\Sigma_\lambda$ one then has 
\begin{equation}
    \Omega_m=\frac{2}{3}\lambda(1+q) \,\,\Rightarrow\,\, \kappa\rho=\lambda Aa^{-3},\,\,\kappa\rho_{\rm DE}=(1-\lambda)Aa^{-3}+B
\end{equation}
Moreover, the statefinder relation \eqref{q} gives, on the leaf $\Sigma_{\lambda}$,
\begin{equation}
   w=\frac{2q-1}{3-2\lambda(1+q)}=-\frac{Ba^3}{(1-\lambda)A+Ba^3}\,. 
\end{equation}
Thus, the leaf $\Sigma_1$ with $\lambda=1$ corresponds to $w=-1$, and on this leaf, the entirety of the $~a^{-3}$ contribution actually comes from nonrelativistic matter. In all other leaves $\Sigma_{\lambda}$ with $0<\lambda<1$, the $~a^{-3}$ contribution to the expansion history comes partially from nonrelativistic matter and partially from the dust-like part of the curvature or nonmetricity fluid.

This one-parameter family of leaves is precisely the geometric phase space representation of the well-known dark degeneracy \cite{Kunz:2009dark}: an expansion history alone cannot tell how its dust-like part is shared between matter and the effective geometric fluid.

\subsection{Why the theory-space/phase space flows look the same for both closures}\label{subsec:why_identical}

In the cosmographic closure approach, as we have formulated in this paper, we use the deceleration parameter $q$ as the clock variable to demonstrate the flow of the cosmological solutions on the base $\{r,m,q\}$ (for $f(R)$) or $\{m,q\}$ (for $f(Q)$). On the other hand, in the equation of state closure approach, we use the matter density abundance parameter $\Omega_m$ as the clock variable to demonstrate the flow of the cosmological solutions on any given leaf, spanned by $\{r,m,\Omega_m\}$ (for $f(R)$) or $\{m,\Omega_m\}$ (for $f(Q)$). For a given leaf $\Sigma_{\lambda}$, these two clock variables are affinely related:
\begin{equation}
    \Omega_m=\frac{2}{3}\lambda(1+q) \Rightarrow \frac{d\Omega_m}{dq}=\frac{2}{3}\lambda\,.
\end{equation}
Moreover, a leaf is an invariant set, meaning that the trajectories on a given leaf stay on that leaf; no trajectory passes through different leaves. The affine relation between the clock variables $\Omega_m$ and $q$, together with the invariant nature of a leaf, implies that the projected phase portrait on the base of \emph{any} given leaf is essentially a scaled version of the phase portrait on the leaf itself. Fig.~\ref{fig:lifted_fQ} makes it clear for $f(Q)$.

The theory space/phase space portraits that we had obtained in Sec.~\ref{sec:cosm_const_mimicking} are the flow on the leaf $\Sigma_1$, and their projections onto the base are the portraits obtained in Sec.~\ref{sec:LCDM_mimicking}. The argument above explains why they look the same.

\subsection{Lines of fixed points and the two faces of $P_2$}\label{subsec:lines}

To perform a fixed point analysis of the lifted system with the fibre coordinate $\Omega_m$, one needs to append the dynamical systems of Sec.~\ref{sec:LCDM_mimicking} with the dynamical equation for $\Omega_m$: Eq.~\eqref{eq:Om_kinematic}. One can notice that at $q=1/2$, both $dq/dN$ and $d\Omega_m/dN$ vanish identically for any value of $\Omega_m$. Geometrically, this means that every fixed point of the base with $q=1/2$, which includes ${\cal P}_1$ and ${\cal P}_2$ of Table \ref{tab:fR_LCDM_fixed_points} and \ref{tab:fQ_LCDM_fixed_points}, is lifted to an entire \emph{line} of fixed points running along the fibre dimension $\Omega_m$. The Jacobian of the lifted system at those lines of fixed points acquires a zero eigenvalue along the line, while its remaining eigenvalues are independent of $\Omega_m$ and coincide with those of the base. 

On the contrary, at the fixed points with $q=-1$, namely $L_c$, $d\Omega_m/dN=-3\Omega_m$ forces $\Omega_m=0$. Hence, the line of fixed points $L_c$ is not lifted along the fibre dimension $\Omega_m$.

The resulting fixed-point structure is summarized in Table~\ref{tab:lifted_fp}.
\begin{table}[!h]
\centering
\renewcommand{\arraystretch}{1.7}
\begin{tabular}{l c c c c}
\hline\hline
Fixed set & $f(R)$: $(r,m,q,\Omega_m)$ & $f(Q)$: $(m,q,\Omega_m)$ & $\Omega_m/f_X$ & Intersection with $\Sigma_\lambda$ \\
\hline
$\mathcal{P}_1$ & $\left(1,\,0,\,\frac{1}{2},\,\Omega_m\right)$ & $\left(0,\,\frac{1}{2},\,\Omega_m\right)$ & $1$ & $\Omega_m=\lambda$ \\
$\mathcal{P}_2$ & $\left(\frac{7+\sqrt{73}}{12},\,\frac{-5+\sqrt{73}}{12},\,\frac{1}{2},\,\Omega_m\right)$ & $\left(-\frac{1}{2},\,\frac{1}{2},\,\Omega_m\right)$ & $0$ & $\Omega_m=\lambda$ \\
$L_c$ & $(2,\,m,\,-1,\,0)$ & $(m,\,-1,\,0)$ & $0$ & the entire line \\
\hline\hline
\end{tabular}
\caption{Fixed points of the lifted phase space of $\Lambda$CDM-mimicking $f(R)$ and $f(Q)$ cosmology, with $0\leq\Omega_m\leq1$. $\mathcal{P}_1$ and $\mathcal{P}_2$ are lines of fixed points along the fibre, each crossing every leaf $\Sigma_\lambda$ exactly once; $L_c$ lies on every leaf. The fixed points $P_1$, $P_2$ of Tables~\ref{tab:fR_LCDM_fixed_points}, \ref{tab:fQ_LCDM_fixed_points} are the projections of $\mathcal{P}_1$, $\mathcal{P}_2$ onto the base, whereas those of Tables~\ref{tab:fR_wm1_fixed_points}, \ref{tab:fQ_wm1_fixed_points} are their intersections with $\Sigma_1$.}
\label{tab:lifted_fp}
\end{table}
The figure \ref{fig:lifted_fQ} provides a clear visual representation of this fact for $f(Q)$ gravity. 

\paragraph{The line $\mathcal{P}_1$:} Along $\mathcal{P}_1$, $f_X=\Omega_m$. Since $\{r,m\}=\{1,0\}$, $f_X$ is constant and the theory is $f=f_X X$. A leaf $\Sigma_\lambda$ crosses $\mathcal{P}_1$ at $\Omega_m=f_X=\lambda$. Each point of the line $\mathcal{P}_1$ in the lifted phase space is an exact GR-matter-dominated solution with the constant-rescaled gravitational coupling $\kappa/\lambda$. The leaf $\Sigma_1$, which corresponds to $w=-1$, intersects the line ${\cal P}_1$ at the canonically normalized point $f_X=\lambda=1$, which is the $P_1$ of Tables~\ref{tab:fR_wm1_fixed_points} and \ref{tab:fQ_wm1_fixed_points}. 

\paragraph{The line $\mathcal{P}_2$:} Along $\mathcal{P}_2$, $\Omega_m/f_X=0$. An exact solution sitting on the line $\mathcal{P}_2$ with finite $f_X$ can therefore exist only at the lowest point, $\Omega_m=0$, of the line. For $f(R)$ gravity, this end point is a vacuum power-law solution. Indeed, on $\mathcal{P}_2$ the parameters $r$ and $m$ are constants with $m=r-1$, so that $f\propto R^{\,n}$ with $n=\frac{7+\sqrt{73}}{12}$. The vacuum solution $a\propto t^{(1-n)(2n-1)/(n-2)}$ of $R^n$ gravity \cite{Carloni:2005Rn} has the matter-like expansion $a\propto t^{2/3}$ ($q=1/2$) precisely when $6n^2-7n-1=0$, whose positive root is exactly this value of $n$. For $f(Q)$ gravity, the end point is the pure $\sqrt{-Q}$ theory ($r=1/2$, $m=-1/2$), which does not contribute to the coincident-gauge background equations \cite{BeltranJimenez:2019tme,Khyllep:2026pku}, so that the vacuum is compatible with any expansion history, a matter-like one in particular. 

The gravity-decoupled $P_2$ of Sec.~\ref{sec:cosm_const_mimicking} is therefore not missing from the $j=1$ analysis of Sec.~\ref{sec:LCDM_mimicking}. It is the opposite end of the very same line of fixed points ${\cal P}_2$, which the reduced systems \eqref{DS_LCDM_f(R)_1} and \eqref{DS_LCDM_f(Q)} display as a single point because they hide the fibre dimension $\Omega_m$. 

\section{Summary and Outlook}\label{sec:summary}

In this paper we have formalized the dynamical system approach to cosmology in $f(R)$ gravity and in the coincident-gauge branch of $f(Q)$ gravity. Departing from the traditional Hubble-normalized formulation, we constructed a dynamical system phase space spanned by three logically distinct sets of variables: a set of directly observable ``kinematic'' variables -- the cosmographic parameters $q,\,j,\,s,\ldots$, the matter density parameter $\Omega_m$, and the dark energy equation of state $w$ together with its derivatives $w',\,w'',\ldots$ -- and a set of ``dynamical'' variables, the theory-space parameters $r,\,m,\,m_2,\ldots$ forming the $\{m_i\}$-hierarchy, which characterize the shape of the underlying theory function $f$. The first two among the above three sets consist of quantities that are directly constrained from data. We showed that this phase space can be closed, and thereby rendered finite-dimensional, in three equivalent ways: by specifying a theory function $f$, by specifying a cosmographic closure relation $j=j(q)$, or by specifying a dark energy equation of state closure relation $w'=w'(w)$.

We illustrated each of the three closure strategies with a concrete worked example.
\begin{itemize}
    \item \textbf{Theory closure.} For the full Hu-Sawicki $f(R)$ family, we identify a constraint surface in the theory space of the form ${\cal F}(r,m,m_2,m_3)=0$, on which any cosmological solution of the Hu-Sawicki family is bound to reside. A unique autonomous system can be obtained in terms of either the set $\{q,j,r,m,m_2,m_3\}$ or the set $\{w,w',\Omega_m,r,m,m_2,m_3\}$. The existence of the GR-matter-dominated family of fixed points $P_1$ and the family of de-Sitter fixed points $L_c$ is generic; however, numerical investigation revealed that a trajectory connecting them may be very fine-tuned to initial conditions near $P_1$. Similarly, for the model $f(Q)=-2\Lambda+Q+\beta\sqrt{-Q}$ with a generic, non-vanishing cosmological constant, we showed that although $m(r)$ is generically double-valued -- introducing a spurious branching of the phase space -- the hierarchy again truncates cleanly, this time already at the level of $m_2$, yielding an unambiguous two-dimensional phase space with a past-attracting GR-matter-dominated fixed point $P_1$ and a stable de-Sitter attractor line $L_c$ (Sec.~\ref{sec:sqrt_f(Q)}).
    \item \textbf{Cosmographic closure.} Imposing the simplest non-trivial cosmographic closure relation, $j=1$ -- i.e.\ kinematic equivalence with $\Lambda$CDM -- we obtained a three-dimensional system for $f(R)$ and a two-dimensional system for $f(Q)$. For $f(R)$, the GR-matter-dominated fixed point is a saddle rather than an attractor, so that a $\Lambda$CDM-mimicking $f(R)$ cosmology generically deviates from GR even in the asymptotic past -- whereas for $f(Q)$ the GR-matter-dominated fixed point is a genuine past attractor. We also found that the $j=1$-closed $f(Q)$ system is exactly integrable (Sec.~\ref{sec:LCDM_mimicking}).
    \item \textbf{Equation-of-state closure.} Imposing the simplest dark energy equation-of-state closure, $w=-1$ identically -- i.e.\ dynamical equivalence with $\Lambda$CDM -- we found that the phase space structure for both $f(R)$ and $f(Q)$ gravity is similar to what we had obtained for the cosmographic closure case with $j=1$ -- even though the dynamical system looks different -- while the fixed point $P_2$ acquires a different physical interpretation (Sec.~\ref{sec:cosm_const_mimicking}).
\end{itemize}

A deeper investigation into the relation between the last two examples turned out to be instructive for the formalism itself (Sec.~\ref{sec:kin_vs_dyn}). The cosmographic closure, as we have formulated in Sec.~\ref{sec:LCDM_mimicking}, is blind to $\Omega_m$. Its complete phase space is obtained by lifting the reduced system with the dynamical equation for $\Omega_m$. The lifted phase space has a skew-product structure and is foliated by invariant leaves labelled by a quantity $\lambda$, whose physical interpretation is the fraction of the genuine nonrelativistic matter contribution to the $~a^{-3}$ part of the Hubble evolution. We identify the result as a geometric phase space picture of the dark degeneracy. The $w=-1$ phase space is precisely the leaf $\lambda=1$, which is diffeomorphic to the reduced $j=1$ system. This is why the $j=1$ closure and the $w=-1$ closure produce identical theory-space flows and phase portraits. In the lifted phase space, the isolated fixed points of the reduced $j=1$ system and the $w=-1$ system, for both $f(R)$ and $f(Q)$ gravity, turn out to be opposite ends of the respective lines of fixed points.

Since our purpose here has been to introduce the general framework and to demonstrate its applicability, we have deliberately restricted our worked examples to simple and physically transparent cases, in particular $j=1$ and $w=-1$. The framework itself is considerably more general and can be extended along several directions. On the theory-closure side, it can be utilized to study other theory functions of physical interest where the traditional dynamical system formulation fails due to the non-invertibility issue with the auxiliary variables. On the cosmographic-closure side, it can be extended to non-$\Lambda$CDM-like cosmic evolutions specified by a non-trivial relation $j=j(q)$, for example the almost-$\Lambda$CDM parametrizations of \cite{Worsley:2026ijo}. Lastly, on the equation-of-state-closure side, to genuinely evolving parametrizations, such as the CPL form $w'(w)=w-w_{(z\to\infty)}$ already noted in Sec.~\ref{subsec:eos_fR}, or oscillating dark energy models. We reserve all of these directions for future work.

It is also possible to extend the applicability of the present formalism to the perturbation level. At the sub-horizon limit and under the quasi-static approximation, the equation governing the matter density perturbation $\delta_m$ at late times, for both $f(R)$ and $f(Q)$ gravity, can be written in the common form
\begin{equation}
    \frac{d^2 \delta_m}{dN^2} + (1 - q) \frac{d \delta_m}{dN} = \frac{3}{2}\frac{\Omega_m}{f_X}\delta_m \quad \Leftrightarrow \quad \frac{d^2 \delta_m}{dN^2} + \left[\frac{1}{2} - \frac{3}{2}w(1-\Omega_m)\right] \frac{d \delta_m}{dN} = \frac{3}{2}\frac{\Omega_m}{f_X}\delta_m\,,\quad (X=R,Q)
\end{equation}
with $\Omega_m/f_X$ to be read off from the appropriate Friedmann constraint. Equivalently, this can be recast as a first-order differential equation for the growth rate $f=\frac{d\ln\delta_m}{dN}$,
\begin{equation}
    \frac{df}{dN} + f^2 + (1-q)f = \frac{3}{2}\frac{\Omega_m}{f_X} \quad \Leftrightarrow \quad \frac{df}{dN} + f^2 + \left[\frac{1}{2} - \frac{3}{2}w(1-\Omega_m)\right]f = \frac{3}{2}\frac{\Omega_m}{f_X}\,,
\end{equation}
where $\Omega_m/f_X$ can be obtained from the respective Friedmann equation. This connects our dynamical system variables directly to the observationally important quantity $f\sigma_8$. Converting derivatives with respect to $N=\ln a$ into derivatives with respect to the cosmological redshift $z$ is straightforward, and all the dynamical equations of this paper can equally well be cast as first-order differential equations in $z$, which is often more convenient for standard numerical solvers. A complete cosmological study of any given model should, of course, involve both a background and a perturbation analysis; the latter is beyond the scope of the present article and is reserved for future work.

Finally, a considerably less trivial extension is to the so-called non-coincident branches of $f(Q)$ cosmology \cite{Hohmann:2021ast,Paliathanasis:2023hqq,Dutta:2025fqw}, in which the presence of a genuinely dynamical connection function precludes the simple algebraic elimination that underlies much of our $f(Q)$ analysis in Sec.~\ref{sec:f(Q)}. Non-coincident gauges are attracting significant phenomenological attention at present \cite{Yang:2024tkw,Paliathanasis:2026end,Paliathanasis:2025dncq,Murtaza:2025bianchi}, and constructing an analogous formulation for them requires special care; we reserve this problem, too, for future work.

\section*{Acknowledgement}

The author is grateful to various invaluable discussions with Khamphee Karwan, Christian Boehmer, Jibitesh Dutta, Wompherdeiki Khyllep, Daniele Gregoris and Peter Dunsby.

\section*{Declaration of generative AI and AI-assisted technologies in the manuscript preparation process}

During the preparation of this work, the author used Claude Opus 5.5, GPT Astra 6, and Gemini Pro at various points to assist with various calculations, derivations, and plots, which were later verified using Wolfram Mathematica 15. After using these tools, the author reviewed and edited the content as needed and takes full responsibility for the content of the published article.

\appendix

\section{Deriving the theory space constraint surface for Hu-Sawicki}\label{app:HS_constraint_surface}

Introduce the auxiliary variable
\begin{equation}
 h\equiv\frac{Rf_{RR}}{f_R-1}=\frac{2n}{1+C_2R^n}-n-1\,.
 \label{eq:HS_h_def}
\end{equation}
One can calculate that for the generic Hu-Sawicki model,
\begin{equation}
    \frac{rm}{rm-(r-1)h} = \frac{R(1-f_R)}{R-f}\,.
\end{equation}
Solving the above two equations, we get
\begin{equation}\label{n_1}
    n=\frac{2rm}{rm-(r-1)h}-1-h\,.
\end{equation}
For the generic Hu-Sawicki model, one can also calculate the following expressions
\begin{equation}
    \frac{dh}{d\ln R}=h(1+m_2-h)=\frac{1}{2}[(h+1)^2-n^2]\,,
\end{equation}
which gives
\begin{equation}\label{n_2}
   n^2=1+3h^2-2hm_2\,. 
\end{equation}
Eliminating $n$ between \eqref{n_1} and \eqref{n_2} gives a cubic equation in $h$:
\begin{equation}
 P(h)\equiv(m_2+1-h)\big[rm-(r-1)h\big]^2
       -2rm\big[rm-(r-1)(h+1)\big]=0\,.
 \label{eq:HS_cubic}
\end{equation}
Next, we differentiate \eqref{n_2} at fixed $n$ and use the recursion relation $dm_2/d\ln R=m_2(1-m_2+m_3)$. The resulting quadratic equation is
\begin{equation}
 Q(h)\equiv3h^2-(4m_2+3)h+m_2(m_3+2)=0\,.
 \label{eq:HS_quad}
\end{equation}

The same $h$ solves both equations \eqref{eq:HS_cubic} and \eqref{eq:HS_quad}, i.e., it is a root of both the polynomials $P(h)$ and $Q(h)$. This is mathematically ensured by the vanishing of the Sylvester determinant between the polynomials, yielding the condition:
\begin{align}
\mathcal{F}(r,m,m_2,m_3)
={}&
\left[
\begin{aligned}
    &9r^2m^2(2m_2-9) \\
    &+6rm(r-1)\bigl[27+m_2(4m_2-6m_3+3)\bigr] \\
    &+(r-1)^2m_2\bigl[9(2m_2+1)(m_3+1)-2m_2(8m_2+3)\bigr]
\end{aligned}
\right]^2
\nonumber\\
&-\bigl(16m_2^2-12m_2m_3+9\bigr)
\left[
\begin{aligned}
    &-9r^2m^2+6rm(r-1)(m_2+3) \\
    &+(r-1)^2m_2(3m_3+3-4m_2)
\end{aligned}
\right]^2
=0 \,.
\end{align}
This is exactly the constraint hypersurface that appears in \eqref{eq:HS_constraint}.

\section{The Hu-Sawicki parameters $\{n,C_1,C_2\}$ in terms of the phase space coordinates}\label{app:HS_parameter_determination}

The auxiliary variable $h$ defined in Eq.\eqref{eq:HS_h_def} can be expressed as
\begin{equation}
h =
\frac{
m_2(r-1)(m_3+2)\bigl[m_2(r-1)-6rm\bigr]
+9rm\bigl[rm(m_2-1)+2(r-1)\bigr]
}{
9r^2m^2-6rm(r-1)(m_2+3)
+(r-1)^2m_2(4m_2-3m_3-3)
}\,.
\end{equation}
Defining $E\equiv H/H_0$, one can evaluate
\begin{subequations}\label{eq:dynvars_to_HSparams}
\begin{align}
n &= \sqrt{1+3h^2-2hm_2}\,,\\[4pt]
C_1H_0^{2n-2}
&= \frac{4mn\,E^{2-2n}}
{(m-h)(n+1+h)^2\,[6(1-q)]^{n-1}}\,,\\[4pt]
C_2 H_0^{2n}
&= \frac{E^{-2n}(n-1-h)}
{(n+1+h)\,[6(1-q)]^n}\,,
\end{align}
Or
\begin{align}
n &= \sqrt{1+3h^2-2hm_2}\,,\\[4pt]
C_1H_0^{2n-2}
&= \frac{4mn\,E^{2-2n}}
{(m-h)(n+1+h)^2\,\bigl[3(1-3w(1-\Omega_m))\bigr]^{n-1}}\,,\\[4pt]
C_2H^{2n}
&= \frac{(n-1-h)E^{-2n}}
{(n+1+h)\,\bigl[3(1-3w(1-\Omega_m))\bigr]^n}\,.
\end{align}
\end{subequations}

\addcontentsline{toc}{section}{References}
\bibliographystyle{unsrt}
\bibliography{refs}

\end{document}